\newif\iffulledition 
\fulleditionfalse 

\documentclass[journal,twoside]{IEEEtran}

\usepackage[utf8]{inputenc}

\usepackage{amsmath,amssymb,amsfonts}

\usepackage{booktabs}
\usepackage{multirow}
\usepackage{threeparttable}
\usepackage{tabularx}
\usepackage{array}
\newcolumntype{M}[1]{>{\centering\arraybackslash}m{#1}}
\newcolumntype{Y}{>{\centering\arraybackslash}X}
\usepackage{makecell}

\usepackage{xurl}
\usepackage{hyperref}

\usepackage[normalem]{ulem}
\usepackage{bbding}

\usepackage{graphicx}
\graphicspath{ {figures/} }
\usepackage{subcaption}
\usepackage{pgfplots}
\usetikzlibrary{arrows, shapes.geometric}
\pgfplotsset{compat=1.5}

\begin{document}
	
\newcommand\ModelName{Semantically Enhanced PCFG}
\newcommand\ModelAcronym{SE\#PCFG}
\newcommand\CrackerName{Semantically Enhanced Password Cracking Architecture}
\newcommand\CrackerAcronym{SEPCA}

\title{\ModelAcronym: \ModelName\ for Password Analysis and Cracking}%
\markboth{IEEE Transactions on Dependable and Secure Computing, Vol.~22, No.~X, Xxxx 2025, doi:\href{https://doi.org/10.1109/TDSC.2025.3547773}{10.1109/TDSC.2025.3547773}}{Wang \MakeLowercase{\textit{et al.}}: \ModelAcronym: \ModelName\ for Password Analysis and Cracking}

\author{Yangde Wang, Weidong Qiu, Peng Tang, Hao Tian and Shujun Li,~\IEEEmembership{Senior Member,~IEEE}%
\thanks{This work was supported by the National Key Research and Development Program of China under grant number 2023YFB3106501.}%
\IEEEcompsocitemizethanks{\IEEEcompsocthanksitem Yangde Wang, Weidong Qiu and Peng Tang are with Shanghai Jiao Tong University, Shanghai, 200240, China.%
\IEEEcompsocthanksitem Shujun Li is with University of Kent, Canterbury, CT2 7NP, UK.%
\IEEEcompsocthanksitem Hao Tian is with Haitong Securities, Shanghai, 200001, China.}%
\thanks{Corresponding co-authors: Yangde Wang (softds@163.com) and Shujun Li (hooklee@gmail.com).}%
}

%


	

\maketitle

\begin{abstract}
Much research has been done on user-generated textual passwords. Surprisingly, semantic information in such passwords remain under-investigated, with passwords created by English- and/or Chinese-speaking users being more studied with limited semantics. This paper fills this gap by proposing a \emph{general framework} based on \emph{semantically enhanced} PCFG (probabilistic context-free grammars) named \ModelAcronym. It allowed us to consider 43 types of semantic information, the richest set considered so far, for password analysis. Applying \ModelAcronym\ to 17 large leaked password databases of user speaking four languages (English, Chinese, German and French), we demonstrate its usefulness and report a wide range of new insights about password semantics at different levels such as cross-website password correlations. Furthermore, based on \ModelAcronym\ and a new systematic smoothing method, we proposed the \CrackerName\ (\CrackerAcronym), and compared its performance against three SOTA (state-of-the-art) benchmarks in terms of the password coverage rate: two other PCFG variants and neural network. Our experimental results showed that \CrackerAcronym\ outperformed all the three benchmarks consistently and significantly across 52 test cases, by up to 21.53\%, 52.55\% and 7.86\%, respectively, at the user-level (with duplicate passwords). At the level of unique passwords, \CrackerAcronym\ also beats the three counterparts by up to 43.83\%, 94.11\% and 11.16\%, respectively.
\end{abstract}


\section{Introduction}
\label{sec:Introduction}

Textual passwords have dominated user authentication on computer systems and the Internet for decades~\cite{Bonneau-SP2012-thequest}. Although many new user authentication methods (e.g., fingerprint and face recognition based methods) have been proposed and used widely on smartphones~\cite{Guo-Wechsler-MobileBiometricsBook2017}, textual passwords remain the most widely used method because none of the new methods can provide a better balance between security and usability. Many people believe that the situation will not change in the foreseeable future~\cite{Herley-SPM2012}.

Trade-offs between security and usability have been well known in the cyber security field~\cite{Lipford-Garfinkel-UsableSecurityBook2014}. For textual passwords, it has been well recognized that users often define easy-to-remember passwords that are not strong enough against password cracking and prefer relying on themselves than using auxiliary tools~\cite{Lyastani-USENIX-Security2018}.

The continuous dominance of textual passwords in user authentication means that it remains important to further our understanding of user-generated passwords to improve password security. There has been quite some research looking into semantic patterns of user-generated passwords, but most of which focused on English-speaking users~\cite{Riddle-CS1989, Brown-ACP2004, Herley-SPM2012, Veras-VizSec2012, Veras-NDSS2014} or more recently Chinese-speaking users~\cite{Li-USENIX-Security14, Wang-CCS2016, Wang-USENIX-Security2019, Shadow-attack-Han-TDSC2018}. However, research covering users speaking more than English and Chinese is still very limited. In addition, past work studied semantic information using stand-alone methods, which means a gap on more reconfigurable frameworks that allow easy incorporation of a rich set of semantic elements. Yet another gap we noticed is that little work has quantitatively analyzed cross-site semantic correlations. Last but not the least, as mentioned in~\cite{TransPCFG2021}, little work has considered applying smoothing techniques to consider unobserved but still plausible passwords to make password cracking methods more generalizable.

This paper fills these gaps via the following main contributions. First, we propose \ModelAcronym, semantically enhanced PCFG, a general framework for analyzing semantics of user-generated passwords. We implemented a prototype of \ModelAcronym\ covering 43 types of password semantic information, the richest set considered so far for password analysis (to the best of our knowledge), including semantic information in four different languages (English, Chinese, German and French), entries in Wikipedia, Wiktionary and Urban Dictionary. Second, by applying our implementation of \ModelAcronym\ to 17 large leaked password databases, we demonstrate its usefulness and report a range of new insights about password semantics and the underlying user behaviors such as cross-website password correlations. Third, we propose \CrackerName\ (\CrackerAcronym), which can leverage training set more effectively, enhanced by a general and systematic smoothing algorithm. Using 52 test cases based on the same 17 password databases (each of four selected databases as the training set and each of the other 13 as the target set), we conducted experiments by comparing the performance of \CrackerAcronym\ against three state-of-the-art password cracking methods in terms of coverage rate: two other variants of the PCFG family -- Weir et al.'s latest implementation~\cite{pcfgcracker} of the original PCFG-based method~\cite{Weir-SP2009-PCFG} and Veras et al.'s method based on their ``Semantic PCFG''~\cite{Veras-NDSS2014} -- and FLA (Fast, Lean, and Accurate) that is n-gram-based and not semantically aware~\cite{Melicher-USENIX-Security2016}. Our experimental results showed that \CrackerAcronym\ outperformed the two other PCFG variants consistently and significantly at both user- and password-levels. With $5 \times 10^9$ guessed passwords, \CrackerAcronym\ performed the best and the average performance across the 52 test cases was improved by up to 21.53\% (user-level), 43.83\% (password-level) for one and 52.55\% (user-level), 94.11\% (password-level) for the other. \CrackerAcronym\ also outperformed FLA by up to 7.86\% (user-level) and 11.16\% (password-level) averagely.

The rest of the paper is organized as follows. The next section summarizes the content of past related studies and provides a detailed comparison with this paper. The third section introduces \ModelAcronym\ and how our implementation was applied to the 17 leaked password databases for password analysis. Section~\ref{sec:PasswordCracking} describes \CrackerAcronym\ in detail and reports experimental results. The fifth section provides an in-depth discussion of the experimental results presented in the previous section, summarizing the implications and guidance offered by our findings for both end users and researchers. Section~\ref{section:Ethics}  includes the ethical statement of this paper, and the last section concludes our work.

\section{Related Work and Comparison}
\label{sec:RelatedWork}

\subsection{Related Work}
\label{subsec:RelatedWork}

\subsubsection{Password modeling methods}
\label{subsubsec:PasswordModelingMethods}

To better understand the habits of people building passwords, plenty of methods were proposed and evaluated by generating guessing passwords. In 2005, Narayanan and Shmatikov~\cite{Narayanan-CCS2005} proposed to use Markov models to guess passwords. Their work was further optimized by Dürmuth et al.\ in 2015~\cite{Durmuth-ESSoS2015} by applying sorting algorithm. In 2014, Ma et al.\ considered varied orders of n-gram Markov models with additive smoothing for cracking English passwords \cite{Ma-SP2014}. In 2016, Melicher et al.~\cite{Melicher-USENIX-Security2016} proposed to use neural networks to model passwords, and they showed that their work could perform well with less memory requirements. Since then, more machine learning-based methods were proposed. These methods not only based on static models learned from training data, but also can follow a more dynamic training process. In 2019, Hitaj et al.~\cite{Hitaj-ACNS2019} proposed to use GAN (Generative Adversarial Networks) to train a password cracker. In 2021, Pasquini et al.~\cite{Pasquini-S&P-2021} showed how the real distribution of target passwords can be interactively learned to facilitate password cracking. In 2023, Wang et al.~\cite{Random-Forest-Wang-US2023} proposed re-encoding the password characters, which makes it possible to use traditional machine learning techniques such as random forests for cracking passwords. In the same year, Xu et al.~\cite{Bi-directional-Xu-US2023} explored and optimized template-based password generation using the bi-transformer technology. In 2024, Li et al.~\cite{passtsl-acisp2024} demonstrated the good generalization ability of pre-training and fine-tuning techniques in password analysis through a two-stage learning process based on transformers.

In 2009, a method based on the so-called probabilistic context-free grammars (PCFG) that can learn higher-level structural patterns than these character-level models, was proposed by Weir et al.~\cite{Weir-SP2009-PCFG}. Their method segment training passwords based on three different types of characters and generate guesses according to probability orders. In 2014, Ma et al.~\cite{Ma-SP2014} reported that PCFG-based methods under-performed whole-string Markov models, revealing that simple PCFGs are not as powerful as they looked. Besides, to our surprise, few prior work aim to optimize PCFG by applying smoothing method. In 2015, Houshmand et al.\ showed the effectiveness of injecting keyboard patterns and using smoothing method limited on them~\cite{Next_Gen_2015}. In 2016, Komanduri designed a smoothing method in an ad hoc way that all types of non-terminals having one pre-defined value~\cite{Komanduri-PhD-Thesis2016}. These two literature can be seen as initial attempts to use smoothing algorithm to optimize PCFG-based methods.

\subsubsection{Password semantic analysis}
\label{subsubsec:PasswordSemanticAnalysis}

Some researchers studied password semantics in order to better understand how users define passwords and to overcome limitations of Markov models and PCFG-based methods. In 1989, Riddle et al.~\cite{Riddle-CS1989} reported that names and dates (especially birthday dates) were often used in user-generated passwords. Through a survey of 218 participants and 1,783 passwords, Brown et al.~\cite{Brown-ACP2004} observed similar phenomena in 2004.

In 2014, Veras et al.~\cite{Veras-NDSS2014} proposed to use NLP techniques to analyze linguistic semantics in user-defined passwords. In 2021, they reported some extended password semantic analysis work in~\cite{Semantic-Analysis-2021}, under the name ``Semantic PCFG''.

Work introduced above mainly considered passwords of English-speaking users. To fill the gap, a number of recent studies looked at leaked passwords from Chinese websites. In 2014, Li et al.~\cite{Li-USENIX-Security14} reported that Chinese users preferred using Pinyin and dates in their passwords. In 2016, Han et al.~\cite{Han-TIFS2016} reported some behavioral differences between Chinese and non-Chinese users on password composition, e.g., Chinese users preferred using digits more but non-Chinese users preferred using letters especially lower-case ones more. At the same year, Wang et al.\ designed a framework named ``TarGuess'' trying to inject various types of personal information to PCFG model to attack specific person over online-attack scenario. In 2017, Wang et al.~\cite{Wang-ASIACCS-2017} reported the observed use of other semantic elements including dates, palindrome, and math. In 2019, Wang et al.~\cite{Wang-USENIX-Security2019} re-confirmed some important semantic elements used by Chinese users such as Pinyin and dates. In 2021, Zhang et al.~\cite{Zhang-ACSAC-2021} looked at how digits in two groups and 12 types were used by Chinese users for defining passwords.

In addition to work on password semantics in English and Chinese passwords, some researchers also looked at passwords defined by users speaking other languages. For instance, AlSabah et al.~\cite{AlSabah2018password_culture} studied semantics in less than 66k passwords and demographic information of users leaked from a Middle Eastern bank, representing diverse cultural backgrounds (Arab, Filipino, Indian, and Pakistani) and non-English/Chinese languages the affected users likely spoke. The semantic information they looked at include names, keyboard patterns, phone numbers and birth years.

On the other hand, some literature also focused on the frequently-used non-linguistic semantics. In 2017, Wang et al.~\cite{Empirical-Rules-2017} studied eight types of transformation rules people usually applied to their passwords. In 2019, Liu et al.~\cite{Reasoning-Liu-SP2019} systematically studied how to identify, order, and apply mangled-rules to widely used cracking tools. In 2021, Xu et al.~\cite{Chunk-2021-CCS} trained a Byte-Pair-Encoding algorithm to automatically obtain chunk vocabularies, and leveraged these information to optimize password models. In 2023, Li et al.~\cite{Mangling-rules-Li-TDSC2023} built an automatic mangling rule generator using density-based clustering to help generating passwords.

Besides research work, there are also many password cracking software tools such as hashcat~\cite{hashcat-forum} and John the Ripper (JtR)~\cite{jtr}. These tools typically use one or more password dictionaries and/or mangling rules to form different attacks, and usually do not incorporate more advanced methods discussed in the research literature. Since such tools are less advanced (in modeling passwords), in the rest of the paper, we will focus on the password analysis and cracking methods reported in research papers only.

\subsection{Comparison With Related Work}
\label{subsec:ComparisonWithRelatedWork}

In this subsection, we explain how our work compare with closest related work. Some terms proposed in our work, especially semantic factors (SFs) and semantic factor types (SFTs), are explained more in Section~\ref{sec:the_proposed_framework}.

Methods based on $n$-grams, such as those proposed by Narayanan \& Shmatikov~\cite{Narayanan-CCS2005}, Melicher et al.~\cite{Melicher-USENIX-Security2016} and Pasquini et al.'~\cite{Pasquini-S&P-2021}, treat each sequence of $n$ consecutive characters as an atomic element for password analysis and cracking, which often cannot be mapped to semantic information in any explicit way. While such methods have been proven very powerful in password cracking (more so than PCFG-based methods), they cannot be used to study password semantics.

Weir et al.'s work~\cite{Weir-SP2009-PCFG, pcfgcracker} started to treat passwords as a series of meaningful components based on character types. Obviously, the lack of semantic awareness in their initial design limits their performance. All previous extensions of Weir et al.'s work were aware of this issue and tried to improve by injecting more semantics. Veras et al.'s work~\cite{Veras-NDSS2014, Semantic-Analysis-2021} introduced NLP tools to develop semantics in English-speaking users. During the same period, \cite{Li-USENIX-Security14, Wang-CCS2016, Wang-zipf-2017, Wang-USENIX-Security2019} tried to better model Chinese behaviors over both online and offline attacking scenario.

Compared with previous work, our work has significant differences in the following key technical aspects. 1) Other work utilized very limited semantic types for password analysis and used ad hoc methods to extract such semantic types, making it difficult to integrate all the different semantic types and extraction methods together into a more comprehensive and expandable framework. For example, Veras et al.~\cite{Veras-NDSS2014, Semantic-Analysis-2021} focused on linguistic semantics in passwords, while others~\cite{Li-USENIX-Security14, Wang-CCS2016, Wang-zipf-2017, Wang-USENIX-Security2019} paid more attention on Chinese names or words in Pinyin. Besides, their choices on the semantic information can be seen as the results of casual observations and appear to be less systematic. In contrast, we followed a more systematic approach to identify different types of semantic information used for password generation by using Google to search for articles about ``How to create strong passwords'', leading to the most comprehensive coverage of semantic types used so far (see~\ref{subsubsec:SFandSFT} for a detailed comparison). 2) Based on the comprehensive semantic information considered in our work, we further propose a new and general smoothing method to address unobserved semantic patterns in passwords, as described in Section~\ref{subsec:SPCA}. 3) To validate the effectiveness of our work, we conducted experiments on the largest collection of leaked password database used in the research literature so far (to the best of our knowledge), which includes passwords from 17 datasets, covering four mainstream languages and 310 million passwords.

\section{\ModelAcronym\ and Password Semantic Analysis}
\label{sec:the_proposed_framework}

In this section, we first describe the conceptual model behind \ModelAcronym, then introduce a streamlined computational process which can tackle different languages and richer semantics, and finally report some selected experimental results by applying our work to analyze 17 large leaked password databases shown in detail in Table~\ref{tab:PasswordDatabases}. All these databases are publicly available and selected according to the following two principles: 1) they should represent a significantly large user population (over 1 million passwords for each) and 2) they should have information about password frequencies to allow richer analysis.

\begin{table}[!ht]
\centering
\caption{The 17 breached databases used in our work.}
\label{tab:PasswordDatabases}
\begin{tabular}{*{6}{c}}
\toprule
No. & Database & Dominating Users & Service & Size & Year\\
\midrule
1 & CSDN & Chinese & Pro. & 6,387,785 & 2011\\
2 & Tianya & Chinese & Soc. & 30,274,001 & 2011\\
3 & 7K7K & Chinese & Ent. & 8,460,641 & 2011\\
4 & 17173 & Chinese & Ent. & 17,942,621 & 2011\\
5 & 178 & Chinese & Ent. & 9,072,688 & 2011\\
6 & Dodonew & Chinese & Pro. & 14,122,756 & 2011\\
7 & Twitter & English & Soc. & 67,095,263 & 2016\\
8 & Webhost & English & Pro. & 14,436,531 & 2015\\
9 & RockYou & English & Soc. & 28,705,927 & 2009\\
10 & MyHeritage & English & Life & 84,825,745 & 2017\\
11 & Gmail & English & Life & 4,663,677 & 2014\\
12 & 8Fit & Germany & Life & 1,121,536 & 2018\\
13 & Eyeem & Germany & Pro. & 4,043,116 & 2018\\
14 & Ge\_Mix1 & Germany & Mix & 6,761,255 & 2018\\
15 & Fr\_Mix1 & French & Mix & 1,302,365 & 2018\\
16 & Fr\_Mix2 & French & Mix & 1,098,418 & 2018\\
17 & Fr\_Mix3 & French & Mix & 10,284,538 & 2018\\
\bottomrule
\end{tabular}
\end{table}

\subsection{Conceptual Model of \ModelAcronym}
\label{subsec:conceptual_model_of_sepcfg}

\subsubsection{Four Structural Levels}

First, we define four password structural levels to better guide analysis of password semantics.

\textbf{1) Characters}: At this level, each character bears the lowest-level information about a password.

\textbf{2) Semantic Factors} (\textbf{word}-level semantics): This level is about a number of consecutive characters (i.e., a word) that together form a semantically meaningful unit, which we call a semantic factor. To indicate what semantic information a semantic factor carries, we call it a \textbf{semantic factor type}. For the sake of brevity, in the following, we use ``SF'' and ``SFT'' to refer to ``semantic factor'' and ``semantic factor type'' respectively. Furthermore, we denote a tuple (SF, \textsc{SFT}) to make it clear what SFT one SF belongs to. 

\textbf{3) Semantic Patterns} (\textbf{password}-level semantics): This level looks at how the whole password is semantically composed of one or more semantic factor types. In the rest of this paper, we use an ordered list of SFTs to denote a password's semantic pattern, e.g., [\textsc{en\_noun}, \textsc{number3}] is the semantic pattern of the password ``\texttt{king123}'', and ``SP'' to refer to ``semantic pattern''.

\textbf{4) Semantic Structure} (\textbf{population}- or \textbf{database}-level semantics): This level is about the overall observable semantic structure of passwords generated by a group of users, reflecting their collective behaviors that could map to one or more shared semantic attributes (e.g., language spoken, age, gender, and website type). For our work, we considered language and website type because they are more available with leaked password databases than other attributes.

Based on the four-level password structure, we can more clearly see how our work differs from others. Specifically, \cite{Narayanan-CCS2005,Melicher-USENIX-Security2016} work more at the first level to build character-to-character transition probabilities without considering any real semantic information. \cite{Veras-NDSS2014,Li-USENIX-Security14,Wang-USENIX-Security2019,Empirical-Rules-2017,Reasoning-Liu-SP2019} explore semantic information at the second level with limit SFTs. In contrast, our work provides a more general way to cover a wide range of semantic information, which can also be tailored for specific password databases. An important contribution of our work is the significant expansion of SFTs covered at the second level to enable much more semantically aware password analysis, as explained in greater detail in the following.

\subsubsection{SFTs and SFs}
\label{subsubsec:SFandSFT}

Understanding the semantic information people use when setting their passwords has never been an easy task because everyone incorporates their life experiences into the password-setting process, resulting in diversity in the semantic components of passwords. In past studies, researches always conducted their analysis through the following steps: manual observation $\to$ classification $\to$ semantic categorisation $\to$ advanced semantic analysis. Unlike past studies, we conducted a systematic search of different types of semantic components considered in previous work and also those mentioned in recommendations for password composition available on the Internet. For the second part, we used ``How to create strong passwords'' as a search query on the Google search engine, and selected the top 10 relevant returned results to identify relevant password composition recommendations. A detailed summary of our results on what we obtained from the 10 websites and what some selected previous studies considered can be found in Table~\ref{tab:advice_on_stronger_passwords}. As can be seen from the table, our work has considered the most comprehensive set of semantic factor types. Note that more SFTs can be easily added by password analysts thanks to the general structure of \ModelAcronym.

\newcommand*\emptycirc[1][0.9ex]{\tikz\draw (0,0) circle (#1);} 
\newcommand*\halfcirc[1][0.9ex]{%
\begin{tikzpicture}
\draw[fill] (0,0)-- (90:#1) arc (90:270:#1) -- cycle ;
\draw (0,0) circle (#1);
\end{tikzpicture}}
\newcommand*\fullcirc[1][0.9ex]{\tikz\fill (0,0) circle (#1);} 

\begin{table*}[!tb]
\renewcommand{\arraystretch}{1.5} 
\centering
\caption{Semantic factor types claimed to be dangerous and how they are considered by previous work and ours.}
\label{tab:advice_on_stronger_passwords}
\begin{threeparttable}
\scriptsize
\begin{tabularx}{\linewidth}{|c|c*{13}{|Y}|}
\hline
\multirow{2}{*}{Source} & \multirow{2}{*}{Word} & \multicolumn{6}{c|}{PI\tnote{a}} & \multicolumn{3}{c|}{SI\tnote{a}} & \multicolumn{3}{c|}{Tricks} & \multirow{2}{*}{Twin\tnote{b}}\\
\cline{3-14}
& & \tiny Name & \tiny Mobile & \tiny Birthday & \tiny Address & \tiny UN\tnote{b} & \tiny Email & \tiny PN$_{1}$\tnote{b} & \tiny ON\tnote{b} & \tiny PN$_{2}$\tnote{b} & \tiny Sequence & \tiny Keyboard & \tiny Sub.\tnote{b} &\\
\hline
Microsoft & \Checkmark & \Checkmark & \XSolidBrush & \XSolidBrush & \XSolidBrush & \XSolidBrush & \XSolidBrush & \Checkmark & \Checkmark & \XSolidBrush & \XSolidBrush & \XSolidBrush & \XSolidBrush & \Checkmark\\
Norton & \Checkmark & \Checkmark & \Checkmark & \Checkmark & \Checkmark & \XSolidBrush & \XSolidBrush & \XSolidBrush & \XSolidBrush & \Checkmark & \XSolidBrush & \XSolidBrush & \XSolidBrush & \Checkmark\\
GCFglobal & \Checkmark & \Checkmark & \XSolidBrush & \XSolidBrush & \XSolidBrush & \Checkmark & \Checkmark & \XSolidBrush & \XSolidBrush & \XSolidBrush & \XSolidBrush & \XSolidBrush & \XSolidBrush & \Checkmark\\
UC Santa & \Checkmark & \Checkmark & \XSolidBrush & \Checkmark & \XSolidBrush & \XSolidBrush & \XSolidBrush & \Checkmark & \XSolidBrush & \XSolidBrush & \XSolidBrush & \XSolidBrush & \XSolidBrush & \XSolidBrush\\
Google & \Checkmark & \Checkmark & \Checkmark & \Checkmark & \Checkmark & \XSolidBrush & \XSolidBrush & \XSolidBrush & \XSolidBrush & \XSolidBrush & \Checkmark & \Checkmark & \XSolidBrush & \Checkmark\\
CMU & \Checkmark & \Checkmark & \Checkmark & \Checkmark & \Checkmark & \XSolidBrush & \XSolidBrush & \Checkmark & \XSolidBrush & \XSolidBrush & \XSolidBrush & \XSolidBrush & \Checkmark & \XSolidBrush\\
AVAST & \Checkmark & \Checkmark & \XSolidBrush & \Checkmark & \Checkmark & \Checkmark & \XSolidBrush & \XSolidBrush & \XSolidBrush & \XSolidBrush & \Checkmark & \XSolidBrush & \Checkmark & \XSolidBrush\\
CISA & \Checkmark & \XSolidBrush & \Checkmark & \Checkmark & \Checkmark & \XSolidBrush & \XSolidBrush & \XSolidBrush & \XSolidBrush & \XSolidBrush & \XSolidBrush & \XSolidBrush & \XSolidBrush & \XSolidBrush\\
WebRoot & \Checkmark & \Checkmark & \XSolidBrush & \Checkmark & \XSolidBrush & \XSolidBrush & \XSolidBrush & \XSolidBrush & \XSolidBrush & \XSolidBrush & \XSolidBrush & \XSolidBrush & \Checkmark & \XSolidBrush\\
Harvard & \Checkmark & \Checkmark & \Checkmark & \Checkmark & \XSolidBrush & \XSolidBrush & \Checkmark & \XSolidBrush & \XSolidBrush & \XSolidBrush & \Checkmark & \Checkmark & \XSolidBrush & \XSolidBrush\\
\hline
Sum & 10 & 9 & 5 & 8 & 5 & 2 & 2 & 3 & 1 & 1 & 3 & 2 & 3 & 4\\
\hline
\cite{Veras-NDSS2014}\tnote{c} & \halfcirc\tnote{d} & \halfcirc & \emptycirc & \emptycirc & \fullcirc & - & \emptycirc & \halfcirc & \halfcirc & \halfcirc & \emptycirc & \emptycirc & \emptycirc & -\\
\cite{Li-USENIX-Security14}\tnote{c} & \halfcirc & \halfcirc & \emptycirc & \fullcirc & \emptycirc & - & \emptycirc & \emptycirc & \emptycirc & \emptycirc & \fullcirc & \fullcirc & \emptycirc & -\\
\cite{Wang-USENIX-Security2019}\tnote{c} & \halfcirc & \fullcirc & \fullcirc & \fullcirc & \fullcirc & - & \emptycirc & \emptycirc & \emptycirc & \emptycirc & \fullcirc & \emptycirc & \emptycirc & -\\
Ours & \fullcirc & \fullcirc & \fullcirc & \fullcirc & \fullcirc & - & \fullcirc & \fullcirc & \fullcirc & \fullcirc & \fullcirc & \fullcirc & \fullcirc & -\\
\hline
\end{tabularx}
\begin{tablenotes}
\item[a] PI and SI are short for ``Personal Information'' and ``Social Information'' respectively.

\item[b] UN, PN$_{1}$, ON, PN$_{2}$ are short for ``User Name'', ``Product Name'', ``Organization Name'' and ``Proper Noun'' respectively. ``Twin'' means passwords used in different websites but from the same user. ``Sub.'' is short for ``Substitution''. Note that password analysis generally clear breached user information to protect privacy, our work exclude ``User Name'' just as other work did.

\item[c] We select three past studies mostly related on studying password semantics, and compare them with our work on these mentioned semantic factor types.

\item[d] \emptycirc, \halfcirc, \fullcirc~mean not, partially and fully considered in each work respectively. As pointed out by~\cite{Wang-USENIX-Security2019}, \cite{Veras-NDSS2014} and \cite{Li-USENIX-Security14} left Chinese Pinyin names unexplored. In terms of words in different languages, the previous three works either focused on English words or added Chinese Pinyin, without considering other languages such as German or French. In addition, the authors of \cite{Veras-NDSS2014} advised to optimize their work by supplementing the corpus containing new terms (e.g. company names, slang or proper nouns) which not appeared in their source corpus.
\end{tablenotes}
\end{threeparttable}
\end{table*}

\textbf{Newly added SFTs}: We introduce 14 new SFTs according to some observed gaps (e.g., what were acknowledged in \cite{Veras-NDSS2014}): 1) 5 SFTs for German words and 5 for French words; 2) Chinese acronyms; 3) \textsc{wkne} and \textsc{ube} to cover proper nouns and slangs; 4) \textsc{consonant} to cover consecutive consonants.

\begin{table*}[!tb]
\renewcommand{\arraystretch}{1.1}
\centering
\caption{43 SFTs used in our implementation of \ModelAcronym.}
\label{tab:SFTs}
\begin{threeparttable}
\begin{tabularx}{\linewidth}{rXrX}
\toprule
SFT & Description\tnote{a} & SFT & Description\\
\midrule
\textsc{email}~\cite{Wang-CCS2016} & Email addresses & \textsc{dn}~\cite{Wang-CCS2016} & Domain namess\\
\textsc{py}~\cite{Li-USENIX-Security14} & Pinyin strings of all Chinese character & \textbf{\textsc{consonants}} & Two or more consecutive consonants can cover many acronyms\\
\textsc{sr4, sr5, $\ldots$}~\cite{Empirical-Rules-2017} & Kinds of Combination of small strings  & \textsc{year}~\cite{Wang-CCS2016} & 4-digit years between 1990 and 2100\\
\textsc{pre1, suf1, $\ldots$}~\cite{Empirical-Rules-2017} & prefixes and suffixes & \textsc{yymmdd, $\ldots$}~\cite{Zhang-ACSAC-2021} & 6- and 8-digit dates in different formats\\ 
\textsc{kb4, kb5, $\ldots$}~\cite{Li-USENIX-Security14} & Keyboard patterns with $4,5,\ldots$ characters & \textsc{cn\_mobile}~\cite{Zhang-ACSAC-2021} & 11-digit mobile numbers (used in China)\\
\textsc{en\_}~\cite{Veras-NDSS2014} & 11 POS tags of English: NOUN, VERB, PRON, ADJ, ADV, ADP, CONJ, DET, PRT, X, NUM & \textbf{\textsc{ge\_, fr\_}} & 5 most common POS tags in German (\textsc{ge\_}) and French (\textsc{fr\_}): NOUN, ADJ, ADV, PRON, VERB\\
\textsc{number1, $\ldots$}~\cite{Weir-SP2009-PCFG} & Numbers with $1,2,\ldots$ digits & \textsc{spec1, $\ldots$}~\cite{Weir-SP2009-PCFG} & Consecutive special characters\\
\textsc{location}~\cite{Wang-USENIX-Security2019} & English names of places\tnote{b} & \textbf{\textsc{wkne}} & \uline{W}i\uline{k}i \uline{n}ame \uline{e}ntity~\cite{wikinameentity}\\
\textsc{month}~\cite{Veras-NDSS2014} & English words for 12 months & \textbf{\textsc{ube}} & \uline{U}r\uline{b}an Dictionary \uline{e}ntity~\cite{urbanDictionary}\\
\textsc{name}~\cite{Wang-USENIX-Security2019} & Male and female names\tnote{c} & \textsc{leet}~\cite{Empirical-Rules-2017} & Leet rules described in \ref{subsubsec:post-processing}\\
\textbf{\textsc{cn\_name\_abbr}} & Acronyms of Chinese names\tnote{d} & \textsc{nn} &Unknown semantics\\
\bottomrule
\end{tabularx}
\begin{tablenotes}
\item[a] 14 newly added SFTs are highlighted in bold, while others were introduced in previous work.

\item[b] Extracted from the world (non-Chinese) location databases in the Chinese instant messaging software Tencent QQ and the Geonames~\cite{GeoNamesData} list of cities.

\item[c] Extracted from a database released by the US Social Security Administration (SSA)~\cite{SSA}, based on a 100\% sample of records of Social Security card applications as of March 2019. The database contains information on gender.

\item[d] 3- and 4-letter only; derived from Chinese names in~\cite{CNnameabbr}.
\end{tablenotes}
\end{threeparttable}
\end{table*}

In addition, we also label any other unknown SFTs as \textsc{nn}. To the best of our knowledge, the 43 SFTs form the richest set of password semantic information considered so far, and serve as a good base line for our implementation and experiments. Table~\ref{tab:SFTs} gives more details about the definitions of these SFTs and sources we used.

\subsection{A Streamlined Computational Process}
\label{subsec:a_streamlined_computational_process}

Based on the conceptual model, we propose a following streamlined computational process of \ModelAcronym\ to automate password semantic analysis in a more general way which consists of three steps: pre-processing, identifying SFTs in segments and post-processing.

\begin{table*}[!tb]
\renewcommand{\arraystretch}{1.5} 
\centering
\caption{Five typical passwords to show details of each step in the computational process of \ModelAcronym. ``---'' means the output of the former step will stay the same after this step.}
\label{table:example_password_semantics}
\scriptsize
\begin{tabularx}{\linewidth}{{c}*6{Y}}
\toprule
Password & Step 1 & Step 2a & Step 2b & Step 3 & Result\\
\hline
\texttt{qwertpassword} & [(\texttt{qwert}, \textsc{KB5}), (\texttt{password}, \textsc{L})] & [(\texttt{qwert}, \textsc{KB5}), (\texttt{password}, \textsc{EN\_NOUN})] & --- & --- & [(\texttt{qwert}, \textsc{KB5}), (\texttt{password}, \textsc{EN\_NOUN})]\\

\texttt{qazqazqaz} & [(\texttt{qazqazqaz}, \textsc{SR9})] & --- & --- & --- & [(\texttt{qazqazqaz}, \textsc{SR9})]\\

\texttt{zhangfei1990} & [(\texttt{zhangfei}, \textsc{L}), (\texttt{1990}, \textsc{D})] & [(\texttt{zhang}, \textsc{PY}), (\texttt{fei}, \textsc{PY}), (\texttt{1990}, \textsc{D})] & [(\texttt{zhang}, \textsc{PY}), (\texttt{fei}, \textsc{PY}), (\texttt{1990}, \textsc{Year})] & --- & [(\texttt{zhang}, \textsc{PY}), (\texttt{fei}, \textsc{PY}), (\texttt{1990}, \textsc{Year})]\\

\texttt{Pa\$\$word} & [(\texttt{Pa}, \textsc{L}), (\texttt{\$\$}, \textsc{SPEC2}), (\texttt{word}, \textsc{L})] & --- & --- & [(\texttt{Pa\$\$word}, \textsc{LEET})] & [(\texttt{Pa\$\$word}, \textsc{LEET})]\\

\texttt{ahnung} & [(\texttt{ahnung}, \textsc{L})] & [(\texttt{ahnung}, \textsc{GE\_NOUN})] & --- & --- & [(\texttt{ahnung}, \textsc{GE\_NOUN})]\\
\bottomrule
\end{tabularx}
\end{table*}

We explain each step with greater details below. Table~\ref{table:example_password_semantics} gives five typical examples of how each step works.
	
\subsubsection{Step~1 -- Pre-processing}
\label{subsubsec:Pre-processing}

Almost all NLP tools consider the change of character type (letter, digit, symbol) as a ``split position'' of consecutive words in a given text. This means that they cannot identify SFs with mixed character types such as ``\texttt{1qaz}'' (a keyboard pattern) and ``\texttt{google.com}'' (a domain name). Therefore, such SFs have to be identified before NLP tools are applied in the next step. Three SFTs we consider here are borrowed from Weir et al.'s implementation and several previous work~ \cite{pcfgcracker,Wang-CCS2016,Li-USENIX-Security14}: keyboard patterns with $n$ characters (\textsc{kb$n$}, where $n\geq 4$), domain names (\textsc{dn}) and email addresses (\textsc{email}). In addition, we also considered three other SFTs with mixed character types: prefixes (\textsc{pre}), suffixes (\textsc{suf}) and repeated strings (\textsc{sr}). We defined the above SFTs in relatively simple manner. Others are free to define more complex versions as needed.

For a given password, the pre-processing step tries to search for all possible SFs falling into the six SFTs following a pre-defined precedence order (\textsc{kb$n$} $>$ \textsc{email} $>$ \textsc{dn} $>$ \textsc{sr$n$} $>$ \textsc{pre} = \textsc{suf}). This order is designed following the implementation of original PCFG~\cite{pcfgcracker}, and adding the three new SFTs in the end for those will not make any ambiguities. After all SFs are labeled, any remaining parts of the password are split into L (Letter), D (Digit) and S (Symbol) segments following the mechanism proposed by Weir et al.'s work \cite{Weir-SP2009-PCFG} for further processing. The first row of Table~\ref{table:example_password_semantics} shows how the pre-processing step works for a given password: \texttt{qwertpassword} $\rightarrow$ [(\texttt{qwert}, \textsc{kb5}), (\texttt{password}, \textsc{L})], where the \textsc{kb5} indicates the identified SFT of \texttt{qwert}. The second row illustrates the identification result of the password ``\texttt{qazqazqaz}''. The remaining parts containing L, D, S segments are for further processing.

\subsubsection{Step~2a -- Identifying SFs in L-Segments}
\label{subsubsec:NLPstep}

After pre-processing, the remaining L-segments can be seen as a combination of multiple SFs (e.g., ``\texttt{wonderbread}''), which are highly language-dependent, therefore NLP tools are needed. In this step, we discuss how \ModelAcronym\ leverage a corpus to obtain richer semantics from L-Segments.

In our implementation of \ModelAcronym, we followed Veras et al.'s work~\cite{Veras-NDSS2014} to choose the widely used NLP library NLTK (\url{https://www.nltk.org/}) to identify linguistic SFs and implement a scoring system based on source and reference corpora and $n$-gram frequencies to disambiguate the results of segmentation. The whole process can be split into two sub-steps: i) further segmenting each input L-segment into smaller linguistic elements (e.g., ``\texttt{sunnyboy}'' into ``\texttt{sunny boy}'') and tagging them, and ii) identifying SFs more than English words.

For sub-step i), we first use several corpora with richer semantics to help NLTK identify SFs. First, we chose to use two widely used English corpora ``Brown'' and ``Web Text'' to cover English words. Then we intersected the German dictionary with word frequencies in WorldLex~\cite{Wordlist-With-Frequence} and Wiktionary of German~\cite{DeWikitionary} to get more commonly used German words. The same was done with the French dictionary in WordLex~\cite{Wordlist-With-Frequence} and Fewiktionary~\cite{FrWikitionary} to produce a French corpus. To cover slangs and proper nouns/phrases, Wikipedia (\textsc{wkne}) and Urban Dictionary (\textsc{ube}) were used to produce two more corpora by concatenating entries they cover. Besides, yet another corpus was produced using a number of ad hoc dictionaries to cover other proposed SFTs such as \textsc{location}.

After segmentation is done, NLTK's POS module is used to directly identify SFs belonging to SFTs with a clear linguistic meanings in English displayed in Table~\ref{tab:SFTs}, which start with \textsc{en\_}. To recognize non-English words without relying on a POS tagger, we chose to inject non-English words into the English POS tagging process as dummy NP words, which can be mapped to the following SFTs via simple string matching: German and French SFTs, \textsc{location}, \textsc{month}, \textsc{male\_name} and \textsc{female\_name}, etc. For any unrecognized segments, we labeled them as \textsc{nn}. Rows 1, 3 and 5 of Table~\ref{table:example_password_semantics} show the results after this step.

\subsubsection{Step~2b -- identifying SFs in D- and S-Segments}
\label{subsubsec:Non-NLPstep}

Step~2a identifies SFs in L-segments, so other SFs are processed in this step using non-NLP methods. For S-segments (i.e., those with special characters only), we treat them as a single SFT \textsc{spec$n$} ($n=1, 2, \ldots$). For D-segments (i.e., numbers), they are processed in two further sub-steps. First, if the length is 4, 6, 8 or 11, the segment will be checked against one of the four SFTs: i) 4-digit years (\textsc{year}), ii) 6-digit dates in the format of \textsc{yymmdd} (Chinese style), \textsc{mmddyy} (American style) and \textsc{ddmmyy} (European style), iii) 8-digit dates in the format of \textsc{yyyymmdd}, \textsc{mmddyyyy}, \textsc{ddmmyyyy}, iv) and 11-digit for mobile phone numbers in China (\textsc{cn\_mobile}). Then, if none of the above SFTs are matched, the number is labeled as \textsc{number$n$} ($n=1, 2, \ldots$). Row 3 of Table~\ref{table:example_password_semantics} shows the result after this step.

\subsubsection{Step~3 -- Post-processing}
\label{subsubsec:post-processing}

After previous steps, a password will be split into multiple sequential SFs. However, for passwords that went through leet transformations, we will end up with a larger number of wrong SFs, e.g., ``\texttt{pa\$\$word}'' will lead to three SFs -- ``\texttt{pa}'', ``\texttt{\$\$}'', ``\texttt{word}''. To fix such problems, we introduce a post-processing step to further process NN-SFs and passwords with too many ($>3$ for our implementation) SFs. According to~\cite{Empirical-Rules-2017}, the top ten transformations (0 $\leftrightarrow$ o, 1 $\leftrightarrow$ i, 3 $\leftrightarrow$ e, 4 $\leftrightarrow$ a, 1 $\leftrightarrow$ !, 1 $\leftrightarrow$ l, 5 $\leftrightarrow$ s, @ $\leftrightarrow$ a, 9 $\leftrightarrow$ 6, \$ $\leftrightarrow$ s) can cover 96.6\% leet pairs, so we decided to consider these leet transformations only. Once detected, we label the whole leet-transformed SFs as a single SF of type \textsc{leet}. Note that the main purpose of this step is to refine segmentation results of previous steps, so more optimizations could be applied. Row 4 of Table~\ref{table:example_password_semantics} gives a visual example.

\subsection{Experimental Results}
\label{subsec:PasswordSemanicsResults}

Now we report selected results of applying our implementation of \ModelAcronym\ to study password semantics of the 17 leaked password databases listed in Table~\ref{tab:PasswordDatabases}. 

\textbf{Attributes of databases}: As mentioned before, the 17 databases were selected to cover two main semantic attributes of online services and their users: language (English, Chinese, German, and French), and service type (Social Networks, Entertainment, Profession, and Life). We noticed that users of each database can be from any country all over the world, but we do not have enough information to determine their nationalities and preferences of speaking language(s). So we categorized databases just based on the dominating users the website served. Note that for English databases of large websites, there are likely many users from non-English-speaking countries, so ``English'' should be treated as ``dominated by English''.

\textbf{Data cleaning}: As with~\cite{Ma-SP2014, Wang-USENIX-Security2019, Random-Forest-Wang-US2023}, we cleaned the databases by removing passwords containing symbols beyond 95 printable ASCII characters or longer than 30 characters. We believe this strategy is reasonable as all these websites only take the 95 printable ASCII characters as legal components of their users' passwords.

\textbf{Segmentation results}: \textsc{nn} can be seen as a good indicator of how well the framework worked. The less \textsc{nn} remain, the more meaningful SFTs are identified. Our experimental results in Table~\ref{table:SuccessRateofSegmentation} showed that our implementation of \ModelAcronym\ can identify 85.11\% and 97.14\% passwords across all 17 databases. Based on these learned semantic information, we report some selected observations at three semantic levels (SFs/SFTs, SPs, and semantic structures) below.

\begin{table}[!tb]
\centering
\caption{Segmentation results over all the 17 databases aligned by language.}
\label{table:SuccessRateofSegmentation}
\begin{threeparttable}
\begin{tabular}{c *{7}{c}}
\toprule
CN & SR (\%)\tnote{a} & EN & SR (\%) & GE & SR (\%) & FR & SR (\%)\\
\midrule
1 & 95.31 & 7 & 90.26 & 12 & 94.84 & 15 & 89.72\\
2 & 95.71 & 8 & 89.00 & 13 & 95.04 & 16 & 89.46\\
3 & 94.80 & 9 & 93.33 & 14 & 88.80 & 17 & 89.29\\
4 & 96.88 & 10 & 85.11 & & & & \\
5 & 96.64 & 11 & 90.99 & & & & \\
6 & 97.14 & & & & & & \\
\bottomrule
\end{tabular}
\begin{tablenotes}
\item[a] ``SR'' is short for ``Success Rate'', which means the percentage of all segmentation results that not contain the SFT of \textsc{nn}.
\end{tablenotes}
\end{threeparttable}
\end{table}

\subsubsection{Analysis of SFs and SFTs}

Past studies have shown frequent use of some SFTs in user-generated passwords, such as numbers, names, dates, and different linguistic elements~\cite{Veras-VizSec2012, Veras-NDSS2014, Wang-CCS2016, Wang-USENIX-Security2019, Semantic-Analysis-2021}, but a systematic look at a more diverse set of SFTs (e.g., the 14 new ones in \ModelAcronym) and SFs is still lacking. To make it easier to identify useful patterns, we re-grouped all the SFTs into 21 groups with a closer semantic relationship: all special characters-based SFs into one (\textsc{special}), all name-related SFs into one (\textsc{name}), all date-related SFs into one (\textsc{date}), all numeric SFs with at least 9 digits into one (\textsc{number9+}), all SFs for a specific language into one (\textsc{en\_sft}s, \textsc{ge\_sft}s, and \textsc{fr\_sft}s), SFs identified during pre-processing and post-processing into \textsc{pre\_processing} and \textsc{post\_processing}, respectively.

\begin{figure}[!b]
\centering
\subfloat[6 Chinese databases]{
	\begin{tikzpicture}
\tikzset{
    every pin/.style={rounded corners=1pt,font=\tiny},
}

\begin{axis}[
width=\linewidth,
height=0.45\linewidth,
ylabel={Frequency (\%)},
font=\scriptsize,
ymax=35,
enlargelimits=0.05,
grid=major,
symbolic x coords={pre\_process,CN\_factor,EN\_factor,GE\_factor,FR\_factor,name,location,mobilephone,post\_process,date,number1,number2,number3,number4,number5,number6,number7,number8,number9+,special,NN},
xtick=data,
xticklabels={,,},
legend entries={1,2,3,4,5,6},
legend columns=3,
legend pos=north east,
legend style={font=\tiny},
inner xsep=2pt
]
\addplot [red,mark=+] table [x=SFT, y=percentage] {figures/semantic-analysis/data/distribution-SF-frequecies/CSDN.txt};
\addplot [red,mark=o]table [x=SFT, y=percentage] {figures/semantic-analysis/data/distribution-SF-frequecies/Tianya.txt};
\addplot [red,mark=asterisk]table [x=SFT, y=percentage] {figures/semantic-analysis/data/distribution-SF-frequecies/7K7K.txt};
\addplot [red,mark=x]table [x=SFT, y=percentage] {figures/semantic-analysis/data/distribution-SF-frequecies/17173.txt};
\addplot [red,mark=triangle]table [x=SFT, y=percentage] {figures/semantic-analysis/data/distribution-SF-frequecies/178.txt};
\addplot [red,mark=pentagon*]table [x=SFT, y=percentage] {figures/semantic-analysis/data/distribution-SF-frequecies/Dodonew.txt};

\fill (axis cs:{[normalized]1},17) circle (1pt) coordinate [pin=85 :{16.87\%}];
\fill (axis cs:{[normalized]2},4) circle (1pt) coordinate [pin=85 :{4.18\%}];
\fill (axis cs:{[normalized]10},2) circle (1pt) coordinate [pin=85 :{2.10\%}];
\fill (axis cs:{[normalized]16},8) circle (1pt) coordinate [pin=85 :{7.86\%}];
\fill (axis cs:{[normalized]20},4) circle (1pt) coordinate [pin=90 :{3.92\%}];
\end{axis}
\end{tikzpicture}
}\\
\subfloat[5 English databases]{
	\begin{tikzpicture}
\tikzset{
    every pin/.style={rounded corners=1pt,font=\tiny},
}
\begin{axis}[
width=\linewidth,
height=0.45\linewidth,
ylabel={Frequency (\%)},
font=\scriptsize,
ymax=35,
enlargelimits=0.05,
grid=major,
symbolic x coords={pre\_process,CN\_factor,EN\_factor,GE\_factor,FR\_factor,name,location,mobilephone,post\_process,date,number1,number2,number3,number4,number5,number6,number7,number8,number9+,special,NN},
xtick=data,
xticklabels={,,},
legend entries={7,8,9,10,11},
legend columns=3,
legend pos=north east,
legend style={font=\tiny},
inner xsep=2pt
]
\addplot [brown,mark=star] table [x=SFT, y=percentage] {./figures/semantic-analysis/data/distribution-SF-frequecies/Twitter.txt};
\addplot [brown,mark=diamond]table [x=SFT, y=percentage] {./figures/semantic-analysis/data/distribution-SF-frequecies/Webhost.txt};
\addplot [brown,mark=square]table [x=SFT, y=percentage] {./figures/semantic-analysis/data/distribution-SF-frequecies/Rockyou.txt};
\addplot [brown,mark=diamond*]table [x=SFT, y=percentage] {./figures/semantic-analysis/data/distribution-SF-frequecies/MyHeritage.txt};
\addplot [brown,mark=triangle*]table [x=SFT, y=percentage] {./figures/semantic-analysis/data/distribution-SF-frequecies/Gmail.txt};

\fill (axis cs:{[normalized]1},6) circle (1pt) coordinate [pin=90 :{6.03\%}];
\fill (axis cs:{[normalized]2},17.5) circle (1pt) coordinate [pin=85 :{17.46\%}];
\fill (axis cs:{[normalized]10},7.5) circle (1pt) coordinate [pin=85 :{7.42\%}];
\fill (axis cs:{[normalized]16},1) circle (1pt) coordinate [pin=85 :{1.19\%}];
\fill (axis cs:{[normalized]20},10) circle (1pt) coordinate [pin=90 :{10.26\%}];
\end{axis}
\end{tikzpicture}
}\\
\subfloat[3 Germany databases]{
	\begin{tikzpicture}
\tikzset{
    every pin/.style={rounded corners=1pt,font=\tiny},
}
\begin{axis}[
width=\linewidth,
height=0.45\linewidth,
ylabel={Frequency (\%)},
font=\scriptsize,
ymax=35,
enlargelimits=0.05,
grid=major,
symbolic x coords={\textsc{pre\_process},\textsc{cn\_sft}s,\textsc{en\_sft}s,\textsc{ge\_sft}s,\textsc{fr\_sft}s,\textsc{name},\textsc{location},\textsc{cn\_mobile},\textsc{post\_process},\textsc{date},\textsc{number1},\textsc{number2},\textsc{number3},\textsc{number4},\textsc{number5},\textsc{number6},\textsc{number7},\textsc{number8},\textsc{number9+},\textsc{special},\textsc{nn}},
xtick=data,
xticklabels={,,},
legend entries={12,13,14},
legend columns=3,
every axis legend/.append style={ at={(0.5,0.65)}, anchor=south },
legend pos=north east,
legend style={font=\tiny},
inner xsep=2pt
]
\addplot [blue,mark=10-pointed star]table [x=SFT, y=percentage] {./figures/semantic-analysis/data/distribution-SF-frequecies/8Fit.txt};
\addplot [blue,mark=*]table [x=SFT, y=percentage] {./figures/semantic-analysis/data/distribution-SF-frequecies/Eyeem.txt};
\addplot [blue,mark=otimes]table [x=SFT, y=percentage] {./figures/semantic-analysis/data/distribution-SF-frequecies/DE3.txt};

\fill (axis cs:{[normalized]1},5.5) circle (1pt) coordinate [pin=90 :{5.69\%}];
\fill (axis cs:{[normalized]2},17) circle (1pt) coordinate [pin=85 :{17.34\%}];
\fill (axis cs:{[normalized]10},4) circle (1pt) coordinate [pin=90 :{4.07\%}];
\fill (axis cs:{[normalized]16},1) circle (1pt) coordinate [pin=85 :{0.99\%}];
\fill (axis cs:{[normalized]20},7) circle (1pt) coordinate [pin=90 :{7.10\%}];
\end{axis}
\end{tikzpicture}
}\\
\subfloat[3 French databases]{
	\begin{tikzpicture}
\tikzset{
    every pin/.style={rounded corners=1pt,font=\tiny},
}
\begin{axis}[
width=\linewidth,
height=0.45\linewidth,
ylabel={Frequency (\%)},
font=\scriptsize,
ymax=35,
enlargelimits=0.05,
grid=major,
symbolic x coords={\textsc{pre\_process},\textsc{cn\_sft}s,\textsc{en\_sft}s,\textsc{ge\_sft}s,\textsc{fr\_sft}s,\textsc{name},\textsc{location},\textsc{cn\_mobile},\textsc{post\_process},\textsc{date},\textsc{number1},\textsc{number2},\textsc{number3},\textsc{number4},\textsc{number5},\textsc{number6},\textsc{number7},\textsc{number8},\textsc{number9+},\textsc{special},\textsc{nn}},
xtick=data,
x tick label style={rotate=60,anchor=east,font=\tiny},
legend entries={15,16,17},
legend columns=3,
every axis legend/.append style={ at={(0.5,0.65)}, anchor=south },
legend pos=north east,
legend style={font=\tiny},
inner xsep=2pt
]
\addplot [black,mark=halfsquare*]table [x=SFT, y=percentage] {./figures/semantic-analysis/data/distribution-SF-frequecies/FR0.txt};
\addplot [black,mark=Mercedes star]table [x=SFT, y=percentage] {./figures/semantic-analysis/data/distribution-SF-frequecies/FR1.txt};
\addplot [black,mark=pentagon]table [x=SFT, y=percentage] {./figures/semantic-analysis/data/distribution-SF-frequecies/FR5.txt};

\fill (axis cs:{[normalized]1},6.5) circle (1pt) coordinate [pin=90 :{6.70\%}];
\fill (axis cs:{[normalized]2},15) circle (1pt) coordinate [pin=85 :{14.97\%}];
\fill (axis cs:{[normalized]10},6) circle (1pt) coordinate [pin=85 :{5.84\%}];
\fill (axis cs:{[normalized]16},1.5) circle (1pt) coordinate [pin=85 :{1.44\%}];
\fill (axis cs:{[normalized]20},10.5) circle (1pt) coordinate [pin=90 :{10.50\%}];
\end{axis}
\end{tikzpicture}
}
\caption{Distribution of combined SFTs in the 17 databases. We can see a clear vision that English, German and French databases have similar distribution at SFT-level except for 10 (MyHeritage). Meanwhile, Chinese databases have similar distribution with each other, but quite different from the other databases. All numbers labeled in each figure are on average.}
\label{fig:SFTfrequencies}
\end{figure}
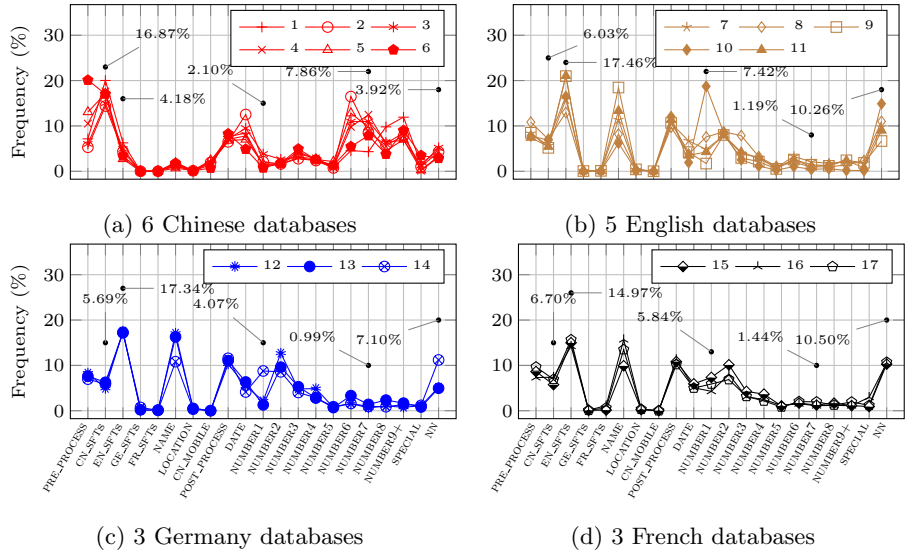

The results led to a number of interesting observations not reported before. \textbf{Preference of languages}: 1) In all databases, Chinese-related SFTs are popular (16.87\%, 1st in Chinese databases, 6.03\%, 7th in English databases, 5.69\%, 6th in German databases and 6.70\%, 6th in French databases), which may be explained that non-Chinese databases are all multi-national so they have a significant number of Chinese-speaking users. 2) For all non-Chinese databases, English SFs as a collective SFT group is either the highest or the second highest. This can be explained by the fact that English is the ``world'' language used widely in all countries. 3) To our surprise, users of German and French databases seemed to prefer English over their native language. Although they have the highest ratio by their own language-related SFTs, but the absolute number is much lower than English-related or even Chinese-related SFTs. This may be explained by non-English-speaking users feeling that using English passwords is more convenient, but more empirical studies involving recruited human participants are needed to understand such a phenomenon more.

\textbf{Numeric SFs}: Past studies~\cite{Li-USENIX-Security14, Wang-USENIX-Security2019, Zhang-ACSAC-2021} have showed the use of numeric segments in user-generated passwords. The richer SFTs used in \ModelAcronym\ still allowed us to observe an interesting new finding: \textbf{Chinese and non-Chinese users had different behaviors} -- Chinese users tended to use longer numeric SFs (with 6-8 digits) than non-Chinese users (with just 1-3 digits).
 
\textbf{Attributions of databases}: No noticeable patterns were observed related to the service type, implying that it may not be a good indicator for analyzing user-generated passwords. In contrast, we can see language plays a key role in the semantic structures at the population/database level: databases sharing the same language have a similar semantic structure, but those labeled with different languages have very different semantic structures. This is a new piece of evidence about users speaking different languages have different password composition behaviors. A visual representation of the above results can be found in Figure~\ref{fig:SFTfrequencies}.

\textbf{New SFTs introduced in \ModelAcronym}: We had interesting observations about the 14 new SFTs described in Section~\ref{subsubsec:SFandSFT}. 1) They play an important role in segmentation results. Averagely 10.38\%, 13.11\%, 12.80\% and 13.47\% passwords consist of these SFTs in Chinese, English, German and French databases, respectively. Out of all these SFTs, \textsc{wkne} is in the majority in all databases, which indicates that this SFT works well in enriching our understanding of password semantics. 2) Some past studies~\cite{Wang-USENIX-Security2019,Semantic-Analysis-2021} reported that in Chinese and English databases, SFs like ``\texttt{love}'' or ``\texttt{ai}'' (the same meaning in Chinese) or ``\texttt{520}x'' (a homophonic number of ``\texttt{I love you}'' in Chinese) appeared very frequently. We observed a similar pattern in German and French databases: ``\texttt{Ich liebe dich}'' and ``\texttt{Je t'aime}'' mean ``I love you'' in German and French, respectively, and they appear from 73--1,329, 64--5,416 times in the language-aligned databases, respectively. On the other hand, we also noticed frequent use of dirty words. For instance, the English phrase ``\texttt{fuckyou}'' appears between 2,110 and 25,357 times in the four English databases. A similar pattern was also observed in Chinese, German and French databases. 3) Some special types of proper nouns/phrases including names of celebrities, large companies/brands and popular games seem much more popular among non-Chinese users than among Chinese users, e.g., ``\texttt{samsung}'' (non-Chinese 0.19\% vs Chinese 0.05\%) and ``\texttt{pokemon}'' (non-Chinese 0.17\% vs Chinese 0.01\%).

\subsubsection{Analysis of SPs}
\label{subsubsec:analysis_of_sps}

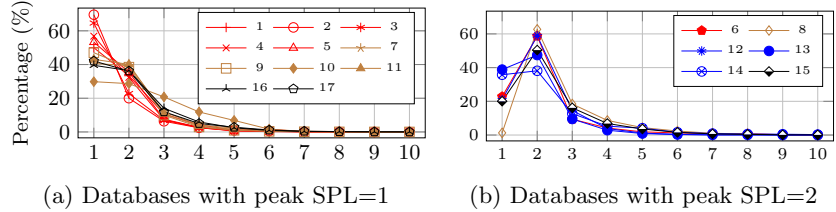
\begin{figure*}[!tb]
\centering
\subfloat[Databases with peak SPL=1]{\begin{tikzpicture}[scale=0.9]
\begin{axis}[
width=0.55\linewidth,
height=3.5cm,
xlabel={SPL},
xtick={1, 2, 3, 4, 5, 6, 7, 8, 9, 10},
ylabel={Percentage (\%)},
font=\scriptsize,
enlargelimits=0.05,
ymax=70,
legend entries={1,2,3,4,5,7,9,10,11,16,17},
legend columns=3,
legend pos=north east,
legend style={font=\tiny},
grid=major
]
\addplot [red,mark=+] table {./figures/semantic-analysis/data/distribution-pattern-length/CSDN.dat};
\addplot [red,mark=o]table {./figures/semantic-analysis/data/distribution-pattern-length/Tianya.dat};
\addplot [red,mark=asterisk]table {./figures/semantic-analysis/data/distribution-pattern-length/7K7K.dat};
\addplot [red,mark=x]table {./figures/semantic-analysis/data/distribution-pattern-length/17173.dat};
\addplot [red,mark=triangle]table {./figures/semantic-analysis/data/distribution-pattern-length/178.dat};
\addplot [brown,mark=star] table {./figures/semantic-analysis/data/distribution-pattern-length/Twitter.dat};
\addplot [brown,mark=square]table {./figures/semantic-analysis/data/distribution-pattern-length/Rockyou.dat};
\addplot [brown,mark=diamond*]table {./figures/semantic-analysis/data/distribution-pattern-length/MyHeritage.dat};
\addplot [brown,mark=triangle*]table {./figures/semantic-analysis/data/distribution-pattern-length/Gmail.dat};
\addplot [black,mark=Mercedes star]table {./figures/semantic-analysis/data/distribution-pattern-length/fr_1.dat};
\addplot [black,mark=pentagon]table {./figures/semantic-analysis/data/distribution-pattern-length/fr_5.dat};
\end{axis}
\end{tikzpicture}}
\subfloat[Databases with peak SPL=2]{\begin{tikzpicture}[scale=0.9]
\begin{axis}[
width=0.55\linewidth,
height=3.5cm,
xlabel={SPL},
xtick={1, 2, 3, 4, 5, 6, 7, 8, 9, 10},
font=\scriptsize ,
enlargelimits=0.05,
ymax=70,
legend entries={6,8,12,13,14,15},
legend columns=2,
legend pos=north east,
legend style={font=\tiny},
grid=major
]
\addplot [red,mark=pentagon*]table {./figures/semantic-analysis/data/distribution-pattern-length/Dodonew.dat};
\addplot [brown,mark=diamond]table {./figures/semantic-analysis/data/distribution-pattern-length/Webhost.dat};
\addplot [blue,mark=10-pointed star]table {./figures/semantic-analysis/data/distribution-pattern-length/8fit.dat};
\addplot [blue,mark=*]table {./figures/semantic-analysis/data/distribution-pattern-length/eyeem.dat};
\addplot [blue,mark=otimes]table {./figures/semantic-analysis/data/distribution-pattern-length/DE_3.dat};
\addplot [black,mark=halfsquare*]table {./figures/semantic-analysis/data/distribution-pattern-length/fr_0.dat};
\end{axis}
\end{tikzpicture}}
\caption{Distributions of SPL in the 17 databases}
\label{fig:LengthDistribution}
\end{figure*}

\textbf{SP length}: For a password, we define the length of SP (SPL) as the number of SFTs included in the SP representing the password. SPL can reflect how complicated a user's mental model was when they generated a password. Figure~\ref{fig:LengthDistribution} shows the results about SPL. In 11 databases (5 Chinese, 4 English and 2 French databases), the majority of SPL is 1, while passwords having SPL of 2 dominate in the other 6 databases. 91.4\% of all passwords have just one to three SFs (39.77\% for SPL = 1, 39.95\% for SPL = 2 and 11.69 for SPL = 3), and almost all (98.3\%) passwords have an SPL no more than five. These results suggest that most users had a relatively simple mental model for generating passwords, which matches the well-reported preference of users for usability over security~\cite{Dell-INFOCOM2010}. Another interesting observation is that the average SPL of all six Chinese databases is 1.702, significantly smaller than that of English (2.136), German (2.037) and French (2.117) databases. Such differences reflect different collective behaviors of Chinese and non-Chinese users. 

\newcommand\drawcolorbox[1]{\fcolorbox{white}{#1}{\rule[1ex]{3ex}{0pt}}}

\definecolor{redlv3}{rgb}{0.99, 0.91, 0.95}
\definecolor{redlv2}{rgb}{0.94, 0.57, 0.75}
\definecolor{redlv1}{rgb}{0.87, 0.12, 0.51}
\definecolor{greenlv3}{rgb}{0.71, 0.88, 0.76}
\definecolor{greenlv2}{rgb}{0.39, 0.75, 0.48}
\definecolor{greenlv1}{rgb}{0.27, 0.64, 0.37}

\subsubsection{Cross-Database Semantic Correlations}
\label{subsubsec:CrossDatabaseCorrelations}

\setlength{\abovedisplayskip}{2pt}
\setlength{\belowdisplayskip}{2pt}

\textbf{Metric to evaluate password semantics at the database level}: The semantic structure of one database can be represented by a discrete probability density (DPD) of each unique SF, SFT or SP. One useful metric capturing similarities and differences of user behaviors is cosine similarity because it is one of the mostly widely used metrics for such purposes~\cite{Han-DM-book2012}. For the three levels, the one at the SFT level will be more robust and easier to calculate because the dimensionality of the DPD at the SFT level is much smaller than that at the other two levels.

It is also possible to define a correlation metric across two or more semantic levels to make the indicator more informative. For instance, assuming that A and B represent the performance forms of a specific SFT on two databases. $A_i$, $B_i$ mean the $i$-th SF in A and B, then we can use the similarity metrics at the SF level for each SFT to adjust the similarity metrics at the SFT level so that the new metric covers both:
\begin{equation}
\text{Sim}_{\mathbf{A},\mathbf{B}}^{\text{SF-SFT}}=\frac{\sum_{i=1}^n(w_{A,B,i}A_iB_i)}{\sqrt{\sum_{i=1}^nA_i^2}\sqrt{\sum_{i=1}^nB_i^2}},
\label{eq:cos_similarity_SF-SFT}
\end{equation}
where $w_{A,B,i}$ is a similarity metric of the SF-level DPDs of the two databases for the $i$-th SFT, with a range of [0,1], and the base-line DPDs are at the SFT level. Similarly, many correlation metrics can be used to calculate $w_{A,B,i}$. In our experiments, we used a simple metric focusing on the average probability of common SFs shared between two databases for a given SFT:
\begin{equation}
w_{A,B,i} = \sum\nolimits_{\text{SF}_j\in\text{SFs}_A\cap\text{SFs}_B} \text{Prob}(\overline{\text{SF}_j}),
\label{eq:similarity_SF-SFTi}
\end{equation}
where $\text{SFs}_A$ and $\text{SFs}_B$ are the sets of all SFs belonging to the $i$-th SFT in Database A and B, respectively, and $\overline{\text{SF}_j}$ is the average occurrence probability of $\text{SF}_j$ in the two databases.

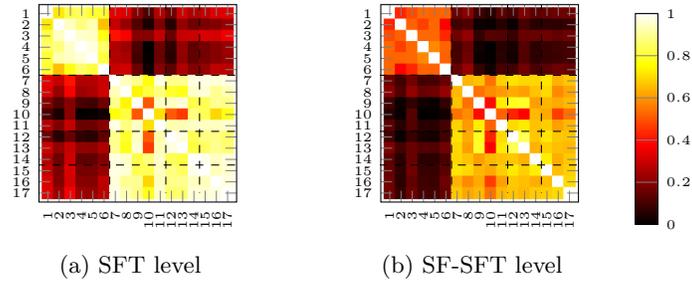
\begin{figure}[!tb]
\centering
\pgfplotsset{
	heatmap axis style/.style={
		width=0.63\linewidth,
		enlargelimits=0.05,
		colorbar=false,
		colormap/hot2,
		ymax=17,
		ymin=1,
		xmax=17,
		xmin=1,
		xtick=data,
		ytick=data,
		x tick label style={rotate=90,font=\tiny},
		y tick label style={font=\tiny},
		axis equal image
	}
}
\subfloat[SFT level]{%
	\begin{tikzpicture}
\begin{axis}[heatmap axis style]
\addplot [
matrix plot,
mesh/cols = 17,
point meta=explicit,
] coordinates {
(1,1) [1](1,2) [0.812047359699153](1,3) [0.880650126301456](1,4) [0.897478275584638](1,5) [0.865122228629134](1,6) [0.865729119677226](1,7) [0.556858485200676](1,8) [0.547115030390958](1,9) [0.462730915257146](1,10) [0.403172649753323](1,11) [0.510313833919382](1,12) [0.438010421741837](1,13) [0.531591148962113](1,14) [0.501389880532664](1,15) [0.517021216859879](1,16) [0.544206624466527](1,17) [0.548707564951662]

(2,1) [0.812047359699153](2,2) [1](2,3) [0.961489452618293](2,4) [0.916401296066746](2,5) [0.948217256967807](2,6) [0.770634907016434](2,7) [0.522507240336467](2,8) [0.478534669379632](2,9) [0.410339346247193](2,10) [0.35647965137117](2,11) [0.454443098352773](2,12) [0.370178951678594](2,13) [0.475711924752289](2,14) [0.437312793214257](2,15) [0.452867987047872](2,16) [0.457829865200114](2,17) [0.482114046780969]

(3,1) [0.880650126301456](3,2) [0.961489452618293](3,3) [1](3,4) [0.968491422807186](3,5) [0.961738891512407](3,6) [0.845171040487877](3,7) [0.596845660849035](3,8) [0.571805254308246](3,9) [0.462314596234209](3,10) [0.463771538977605](3,11) [0.521786700066085](3,12) [0.431995842642755](3,13) [0.525059893803114](3,14) [0.529746997423219](3,15) [0.545282465428738](3,16) [0.5401527551439](3,17) [0.567961363782218]

(4,1) [0.897478275584638](4,2) [0.916401296066746](4,3) [0.968491422807186](4,4) [1](4,5) [0.977548429926836](4,6) [0.902198239226301](4,7) [0.503263721606802](4,8) [0.484773676683575](4,9) [0.401178904742909](4,10) [0.329596413881165](4,11) [0.434115653304213](4,12) [0.374883254457971](4,13) [0.474255973671916](4,14) [0.423398246466687](4,15) [0.446154881266684](4,16) [0.462947243762826](4,17) [0.483182361721917]

(5,1) [0.865122228629134](5,2) [0.948217256967807](5,3) [0.961738891512407](5,4) [0.977548429926836](5,5) [1](5,6) [0.874448076016568](5,7) [0.505766748153546](5,8) [0.495998894559465](5,9) [0.392482622430348](5,10) [0.335910863931126](5,11) [0.433938411142856](5,12) [0.369075484949081](5,13) [0.471389635418442](5,14) [0.426087218920468](5,15) [0.446628000147584](5,16) [0.460966676634112](5,17) [0.479909981705863]

(6,1) [0.865729119677226](6,2) [0.770634907016434](6,3) [0.845171040487877](6,4) [0.902198239226301](6,5) [0.874448076016568](6,6) [1](6,7) [0.528347656385037](6,8) [0.533772274016288](6,9) [0.436778315111013](6,10) [0.347573951288474](6,11) [0.461322314162022](6,12) [0.422053794458728](6,13) [0.502465735123307](6,14) [0.45934879769288](6,15) [0.476828994245694](6,16) [0.516691626165228](6,17) [0.513439367333784]

(7,1) [0.556858485200676](7,2) [0.522507240336467](7,3) [0.596845660849035](7,4) [0.503263721606802](7,5) [0.505766748153546](7,6) [0.528347656385037](7,7) [1](7,8) [0.955079863158894](7,9) [0.924899715431772](7,10) [0.838593082128483](7,11) [0.981459775953339](7,12) [0.923981779726882](7,13) [0.926964507743569](7,14) [0.979687674143133](7,15) [0.982428082892395](7,16) [0.975109302999457](7,17) [0.984000072260324]

(8,1) [0.547115030390958](8,2) [0.478534669379632](8,3) [0.571805254308246](8,4) [0.484773676683575](8,5) [0.495998894559465](8,6) [0.533772274016288](8,7) [0.955079863158894](8,8) [1](8,9) [0.824202713417046](8,10) [0.874469646769607](8,11) [0.916290354443153](8,12) [0.861657946104773](8,13) [0.848048914910703](8,14) [0.965724893139694](8,15) [0.972529962455285](8,16) [0.92776680674009](8,17) [0.950618860820403]

(9,1) [0.462730915257146](9,2) [0.410339346247193](9,3) [0.462314596234209](9,4) [0.401178904742909](9,5) [0.392482622430348](9,6) [0.436778315111013](9,7) [0.924899715431772](9,8) [0.824202713417046](9,9) [1](9,10) [0.667390479910767](9,11) [0.961126175901366](9,12) [0.961669096938224](9,13) [0.974735069039573](9,14) [0.880571382213914](9,15) [0.887949605834476](9,16) [0.92993901847693](9,17) [0.92120416809624]

(10,1) [0.403172649753323](10,2) [0.35647965137117](10,3) [0.463771538977605](10,4) [0.329596413881165](10,5) [0.335910863931126](10,6) [0.347573951288474](10,7) [0.838593082128483](10,8) [0.874469646769607](10,9) [0.667390479910767](10,10) [1](10,11) [0.802416415429122](10,12) [0.670195922494868](10,13) [0.637566152347066](10,14) [0.914060922513832](10,15) [0.888598554610714](10,16) [0.793324111651684](10,17) [0.848059646633217]

(11,1) [0.510313833919382](11,2) [0.454443098352773](11,3) [0.521786700066085](11,4) [0.434115653304213](11,5) [0.433938411142856](11,6) [0.461322314162022](11,7) [0.981459775953339](11,8) [0.916290354443153](11,9) [0.961126175901366](11,10) [0.802416415429122](11,11) [1](11,12) [0.946140261967214](11,13) [0.950325076173431](11,14) [0.963661908567245](11,15) [0.959507646789459](11,16) [0.958983911946699](11,17) [0.966451659517505]

(12,1) [0.438010421741837](12,2) [0.370178951678594](12,3) [0.431995842642755](12,4) [0.374883254457971](12,5) [0.369075484949081](12,6) [0.422053794458728](12,7) [0.923981779726882](12,8) [0.861657946104773](12,9) [0.961669096938224](12,10) [0.670195922494868](12,11) [0.946140261967214](12,12) [1](12,13) [0.978751034686554](12,14) [0.888152764370219](12,15) [0.91800043385933](12,16) [0.92061545148117](12,17) [0.904455474677365]

(13,1) [0.531591148962113](13,2) [0.475711924752289](13,3) [0.525059893803114](13,4) [0.474255973671916](13,5) [0.471389635418442](13,6) [0.502465735123307](13,7) [0.926964507743569](13,8) [0.848048914910703](13,9) [0.974735069039573](13,10) [0.637566152347066](13,11) [0.950325076173431](13,12) [0.978751034686554](13,13) [1](13,14) [0.876360747884305](13,15) [0.897193120091958](13,16) [0.92757633239341](13,17) [0.905599664037521]

(14,1) [0.501389880532664](14,2) [0.437312793214257](14,3) [0.529746997423219](14,4) [0.423398246466687](14,5) [0.426087218920468](14,6) [0.45934879769288](14,7) [0.979687674143133](14,8) [0.965724893139694](14,9) [0.880571382213914](14,10) [0.914060922513832](14,11) [0.963661908567245](14,12) [0.888152764370219](14,13) [0.876360747884305](14,14) [1](14,15) [0.988624609496544](14,16) [0.958051004960876](14,17) [0.975350813313705]

(15,1) [0.517021216859879](15,2) [0.452867987047872](15,3) [0.545282465428738](15,4) [0.446154881266684](15,5) [0.446628000147584](15,6) [0.476828994245694](15,7) [0.982428082892395](15,8) [0.972529962455285](15,9) [0.887949605834476](15,10) [0.888598554610714](15,11) [0.959507646789459](15,12) [0.91800043385933](15,13) [0.897193120091958](15,14) [0.988624609496544](15,15) [1](15,16) [0.954202344078908](15,17) [0.969880918152841]

(16,1) [0.544206624466527](16,2) [0.457829865200114](16,3) [0.5401527551439](16,4) [0.462947243762826](16,5) [0.460966676634112](16,6) [0.516691626165228](16,7) [0.975109302999457](16,8) [0.92776680674009](16,9) [0.92993901847693](16,10) [0.793324111651684](16,11) [0.958983911946699](16,12) [0.92061545148117](16,13) [0.92757633239341](16,14) [0.958051004960876](16,15) [0.954202344078908](16,16) [1](16,17) [0.979407852500847]

(17,1) [0.548707564951662](17,2) [0.482114046780969](17,3) [0.567961363782218](17,4) [0.483182361721917](17,5) [0.479909981705863](17,6) [0.513439367333784](17,7) [0.984000072260324](17,8) [0.950618860820403](17,9) [0.92120416809624](17,10) [0.848059646633217](17,11) [0.966451659517505](17,12) [0.904455474677365](17,13) [0.905599664037521](17,14) [0.975350813313705](17,15) [0.969880918152841](17,16) [0.979407852500847](17,17) [1]
};
\addplot[dashed,domain=0:20] {6.5};
\addplot[dashed,domain=0:20] {11.5};
\addplot[dashed,domain=0:20] {14.5};
\addplot[dashed,domain=1:19] coordinates{
(6.5,0) (6.5,1) (6.5,2)  (6.5,3) (6.5,4) (6.5,5) (6.5,6) (6.5,7)
(6.5,8) (6.5,9) (6.5,10) (6.5,11) (6.5,12) (6.5,13) (6.5,14) (6.5,15) (6.5,16) (6.5,17)
};
\addplot[dashed,domain=1:19] coordinates{
(11.5,0) (11.5,1) (11.5,2)  (11.5,3) (11.5,4) (11.5,5) (11.5,6) (11.5,7) (11.5,8) (11.5,9) (11.5,10) (11.5,11) (11.5,12) (11.5,13) (11.5,14) (11.5,15) (11.5,16) (11.5,17)
};
\addplot[dashed,domain=1:19] coordinates{
(14.5,0) (14.5,1) (14.5,2)  (14.5,3) (14.5,4) (14.5,5) (14.5,6) (14.5,7) (14.5,8) (14.5,9) (14.5,10) (14.5,11) (14.5,12) (14.5,13) (14.5,14) (14.5,15) (14.5,16) (14.5,17)
};

\end{axis}
\end{tikzpicture}
	\label{fig:correlationsOnSFT}
}
\subfloat[SF-SFT level]{%
	\begin{tikzpicture}
\begin{axis}[heatmap axis style]
\addplot [
matrix plot,
mesh/cols = 17,
point meta=explicit,
] coordinates {
(1,1) [1.0](1,2) [0.567088881187654](1,3) [0.610757292933713](1,4) [0.642082430858978](1,5) [0.637944941950465](1,6) [0.640350275236817](1,7) [0.364158824511998](1,8) [0.397367609580649](1,9) [0.315427138878098](1,10) [0.288498887342072](1,11) [0.344316925227585](1,12) [0.318182628840058](1,13) [0.384122547557093](1,14) [0.344747593263147](1,15) [0.355917541004683](1,16) [0.366659486737298](1,17) [0.368400804744366]

(2,1) [0.56708888118766](2,2) [1.0](2,3) [0.634521666876561](2,4) [0.597993690527896](2,5) [0.639720708618181](2,6) [0.525021405285627](2,7) [0.315578873778835](2,8) [0.331036250235651](2,9) [0.258615441158027](2,10) [0.25238882860469](2,11) [0.287601735765197](2,12) [0.256152103206954](2,13) [0.320552569646632](2,14) [0.287106901772085](2,15) [0.295294862934531](2,16) [0.292234189501314](2,17) [0.303539822866085]

(3,1) [0.610757292933713](3,2) [0.634521666876561](3,3) [1.0](3,4) [0.643608414151206](3,5) [0.642430377821615](3,6) [0.570197588011809](3,7) [0.371292242985018](3,8) [0.4058264377308](3,9) [0.298159224871368](3,10) [0.340643652979081](3,11) [0.334743622788453](3,12) [0.307127366403672](3,13) [0.360914844702062](3,14) [0.360038564422188](3,15) [0.366945591945149](3,16) [0.351311323159949](3,17) [0.364876979516916]

(4,1) [0.642082430858978](4,2) [0.597993690527895](4,3) [0.643608414151204](4,4) [1.0](4,5) [0.678864033044784](4,6) [0.63776030406664](4,7) [0.319573590549536](4,8) [0.359294006286353](4,9) [0.264082383086097](4,10) [0.243732731366424](4,11) [0.288565052978837](4,12) [0.271863811161343](4,13) [0.33695989536366](4,14) [0.291364897202156](4,15) [0.306572614361926](4,16) [0.312632705689522](4,17) [0.319776260515543]

(5,1) [0.637944941950465](5,2) [0.639720708618181](5,3) [0.642430377821615](5,4) [0.678864033044783](5,5) [1.0](5,6) [0.634405435682144](5,7) [0.32234114300912](5,8) [0.368760258324846](5,9) [0.256513428612509](5,10) [0.248097070163198](5,11) [0.290527124682636](5,12) [0.265604648038663](5,13) [0.336237661820377](5,14) [0.293102159083348](5,15) [0.306559625423142](5,16) [0.310901273331514](5,17) [0.317756882279645]

(6,1) [0.640350275236819](6,2) [0.525021405285627](6,3) [0.570197588011809](6,4) [0.637760304066645](6,5) [0.634405435682144](6,6) [1.0](6,7) [0.358485568075378](6,8) [0.411004314500263](6,9) [0.30142938830584](6,10) [0.262830733567646](6,11) [0.321934500836151](6,12) [0.315010371187292](6,13) [0.369555610312624](6,14) [0.327115855969082](6,15) [0.344277142548296](6,16) [0.364834146576373](6,17) [0.356337214633325]

(7,1) [0.364158824511997](7,2) [0.315578873778835](7,3) [0.371292242985018](7,4) [0.319573590549537](7,5) [0.322341143009121](7,6) [0.358485568075378](7,7) [1.0](7,8) [0.731012751671578](7,9) [0.744734021310339](7,10) [0.652694612255007](7,11) [0.759514406698682](7,12) [0.779203753557003](7,13) [0.7670222650586](7,14) [0.765620318615857](7,15) [0.782987672023895](7,16) [0.725450414850371](7,17) [0.748905086584588]

(8,1) [0.397367609580649](8,2) [0.331036250235649](8,3) [0.4058264377308](8,4) [0.359294006286353](8,5) [0.368760258324846](8,6) [0.411004314500263](8,7) [0.731012751671575](8,8) [1.0](8,9) [0.653204055847915](8,10) [0.682107305479041](8,11) [0.699156517458522](8,12) [0.719400870751724](8,13) [0.704139491509484](8,14) [0.744895035541134](8,15) [0.768465600505094](8,16) [0.679865887750497](8,17) [0.712962299438814]

(9,1) [0.315427138878098](9,2) [0.258615441158026](9,3) [0.298159224871366](9,4) [0.264082383086098](9,5) [0.256513428612509](9,6) [0.30142938830584](9,7) [0.744734021310343](9,8) [0.653204055847918](9,9) [1.0](9,10) [0.538898965059599](9,11) [0.793311656337786](9,12) [0.833906878828671](9,13) [0.841985535577421](9,14) [0.703834648222122](9,15) [0.729902927117124](9,16) [0.716800445534297](9,17) [0.732099003439046]

(10,1) [0.288498887342072](10,2) [0.25238882860469](10,3) [0.340643652979081](10,4) [0.243732731366424](10,5) [0.248097070163198](10,6) [0.262830733567646](10,7) [0.652694612255007](10,8) [0.682107305479041](10,9) [0.538898965059598](10,10) [1.0](10,11) [0.635263887040741](10,12) [0.564374364012827](10,13) [0.527133711549732](10,14) [0.733655171478297](10,15) [0.716010773976606](10,16) [0.594100516219534](10,17) [0.652102716110779]

(11,1) [0.344316925227584](11,2) [0.287601735765196](11,3) [0.334743622788451](11,4) [0.288565052978837](11,5) [0.290527124682636](11,6) [0.321934500836151](11,7) [0.759514406698675](11,8) [0.699156517458519](11,9) [0.793311656337787](11,10) [0.635263887040743](11,11) [1.0](11,12) [0.769949187008759](11,13) [0.78891073395118](11,14) [0.741881476366148](11,15) [0.757162287818914](11,16) [0.70520759492124](11,17) [0.732230627667871]

(12,1) [0.318182628840058](12,2) [0.256152103206954](12,3) [0.307127366403672](12,4) [0.271863811161343](12,5) [0.265604648038663](12,6) [0.315010371187292](12,7) [0.779203753557002](12,8) [0.719400870751728](12,9) [0.83390687882867](12,10) [0.564374364012826](12,11) [0.769949187008758](12,12) [1.0](12,13) [0.874296356818938](12,14) [0.748773689969654](12,15) [0.787198499829307](12,16) [0.751545488248328](12,17) [0.746564476926621]

(13,1) [0.384122547557093](13,2) [0.320552569646626](13,3) [0.360914844702062](13,4) [0.33695989536366](13,5) [0.336237661820377](13,6) [0.369555610312626](13,7) [0.767022265058599](13,8) [0.704139491509479](13,9) [0.841985535577421](13,10) [0.527133711549732](13,11) [0.788910733951181](13,12) [0.874296356818938](13,13) [1.0](13,14) [0.72655097119917](13,15) [0.758996505333362](13,16) [0.741436367939148](13,17) [0.733893062038428]

(14,1) [0.344747593263147](14,2) [0.287106901772085](14,3) [0.360038564422189](14,4) [0.291364897202155](14,5) [0.293102159083348](14,6) [0.327115855969082](14,7) [0.765620318615859](14,8) [0.74489503554114](14,9) [0.703834648222126](14,10) [0.733655171478295](14,11) [0.741881476366157](14,12) [0.748773689969651](14,13) [0.726550971199174](14,14) [1.0](14,15) [0.795081404475907](14,16) [0.72067049099174](14,17) [0.753246210784579]

(15,1) [0.355917541004684](15,2) [0.295294862934532](15,3) [0.366945591945148](15,4) [0.306572614361926](15,5) [0.306559625423142](15,6) [0.344277142548296](15,7) [0.782987672023897](15,8) [0.768465600505093](15,9) [0.729902927117125](15,10) [0.716010773976608](15,11) [0.757162287818907](15,12) [0.787198499829306](15,13) [0.758996505333363](15,14) [0.795081404475908](15,15) [1.0](15,16) [0.739657219773894](15,17) [0.75818492895858]

(16,1) [0.366659486737299](16,2) [0.292234189501313](16,3) [0.351311323159949](16,4) [0.312632705689522](16,5) [0.310901273331514](16,6) [0.364834146576373](16,7) [0.725450414850366](16,8) [0.679865887750495](16,9) [0.716800445534297](16,10) [0.594100516219534](16,11) [0.705207594921241](16,12) [0.751545488248326](16,13) [0.741436367939154](16,14) [0.720670490991734](16,15) [0.739657219773894](16,16) [1.0](16,17) [0.770179323226771]

(17,1) [0.368400804744365](17,2) [0.303539822866082](17,3) [0.364876979516914](17,4) [0.319776260515543](17,5) [0.317756882279645](17,6) [0.356337214633325](17,7) [0.748905086584582](17,8) [0.712962299438818](17,9) [0.732099003439043](17,10) [0.652102716110777](17,11) [0.732230627667871](17,12) [0.746564476926622](17,13) [0.733893062038429](17,14) [0.753246210784585](17,15) [0.758184928958576](17,16) [0.770179323226766](17,17) [1.0]
};
\addplot[dashed,domain=0:20] {6.5};
\addplot[dashed,domain=0:20] {11.5};
\addplot[dashed,domain=0:20] {14.5};
\addplot[dashed,domain=1:19] coordinates{
	(6.5,0) (6.5,1) (6.5,2)  (6.5,3) (6.5,4) (6.5,5) (6.5,6) (6.5,7)
	(6.5,8) (6.5,9) (6.5,10) (6.5,11) (6.5,12) (6.5,13) (6.5,14) (6.5,15) (6.5,16) (6.5,17)
};
\addplot[dashed,domain=1:19] coordinates{
	(11.5,0) (11.5,1) (11.5,2)  (11.5,3) (11.5,4) (11.5,5) (11.5,6) (11.5,7) (11.5,8) (11.5,9) (11.5,10) (11.5,11) (11.5,12) (11.5,13) (11.5,14) (11.5,15) (11.5,16) (11.5,17)
};
\addplot[dashed,domain=1:19] coordinates{
	(14.5,0) (14.5,1) (14.5,2)  (14.5,3) (14.5,4) (14.5,5) (14.5,6) (14.5,7) (14.5,8) (14.5,9) (14.5,10) (14.5,11) (14.5,12) (14.5,13) (14.5,14) (14.5,15) (14.5,16) (14.5,17)
};
\end{axis}
\end{tikzpicture}
	\label{fig:correlationsOnSF}
}
\begin{tikzpicture}
	\pgfplotscolorbardrawstandalone[ 
	colormap/hot2,
	colorbar/width=3mm,
	point meta min=0,
	point meta max=1,
	colorbar style={
		width=3mm,
		font=\scriptsize,
		height=0.4\linewidth,
	}
	]
\end{tikzpicture}
\caption{Cross-database semantic correlation values at the SFT level and those at the combined SF-SFT level, according to~\cite{Han-DM-book2012} and Eq.~\eqref{eq:cos_similarity_SF-SFT}, respectively. The x- and y-axis show the indices of the 17 databases shown in Table~\ref{tab:PasswordDatabases}.}
\label{fig:Correlation_of_SFs_and_SFTs}
\end{figure}

\textbf{Cross-database semantic correlations at the SFT and SF levels}: Following the equations above, we can calculate the overall semantic correlation between any two given databases at different semantic levels. Figure~\ref{fig:Correlation_of_SFs_and_SFTs} shows the cross-database semantic correlation values between each pair of the 17 databases as a diagonally symmetric heatmap, using~\cite{Han-DM-book2012} and Eqs.~\eqref{eq:cos_similarity_SF-SFT}, respectively. The dashed lines in the heatmaps separate Chinese (1-6), English (7-11), German (12-14) and French (15-17) databases to show the language-dependent patterns more clearly. From the two heatmaps, we can see a number of visual patterns. First, there are two clearly non-overlapping areas -- one for Chinese databases, and the other for non-Chinese databases, indicating that Chinese and non-Chinese users have very different collective behaviors. One possible reason of this pattern is that Chinese websites are more dominated by Chinese-speaking users, but Western websites have a more mixed groups of users who speak different languages. Another reason is that Western languages are linguistically more similar to each other than Chinese to any Western languages. Second, users of Myheritage (10) display very different habits from all the other databases. This phenomenon is echoed later by the poorer password cracking performance against Myheritage using other databases as the training database (see Sections~\ref{subsec:CrackingExperimentalResult}). Finally but equally interestingly, comparing the two heatmaps, the correlation values between Chinese databases drop significantly when SFs are considered to weigh the SFT-level correlations, suggesting that Chinese users share more common behaviors on the selection of SFTs but they behave less similarly on selections of SFs. This phenomenon is much less obvious for non-Chinese databases, suggesting that Western users are more consistent in choosing both SFs and SFTs.

\section{Semantically Enhanced Password Cracking}
\label{sec:PasswordCracking}

Thanks to the enhanced semantic awareness, \ModelAcronym\ clearly has the potential to be used for designing more powerful PCFG-based password cracking methods. Combining \ModelAcronym\ with a systematic model smoothing method, we developed \CrackerName\ (\CrackerAcronym), a new password cracking architecture that was shown to be able to outperform mainstream SOTA password cracking methods under the scene of real-attacking. The main idea of model smoothing is to address SFs that are not present in training sets but appear in the targets. This problem was first mentioned in~\cite{Weir-SP2009-PCFG}. Surprisingly, very few researchers have studied how to practically and systematically smooth a password model, which so far is mainly done by injecting extra information such as new dictionaries with a fixed but ad hoc coefficient.In the following, we first explain the design and implementation of \CrackerAcronym, as one of the technical contribution of our work, which allowing us apply a natural way to assign a set of non-zero probabilities to unobserved SFs, then discuss how we conducted our experiments, and finally present the results comparing with three state-of-the-art methods and new insights learned.

\subsection{The Proposed New Architecture \CrackerAcronym}
\label{subsec:SPCA}

A generic probabilistic context-free grammar \emph{G} can be defined by a quintuple, \emph{G=}(\emph{M, T, R, S, P}), where \emph{M} and \emph{T} represent the set of non-terminal and terminal symbols respectively. \emph{S} is the start symbol belonging to \emph{M}. \emph{R} is the set of production rules and \emph{P} contains the probabilities of each rule in \emph{R}.

In \CrackerAcronym, \emph{T} is the set of all SFs, \emph{M} is the union of \emph{T} and \emph{S}. The production rules can be categorized into two groups: 1) from \emph{S} to a certain SP, following $\sum_{k}P(\emph{S} \rightarrow \text{SP}_k) = 1$. 2) from a SFT to a certain SF, and $\forall i, \sum_{j}P(\text{SFT}_i \rightarrow \text{SF}_j) = 1$. Under the above grammar, we can calculate the probability of any given password as follows to allow ranking passwords for cracking purposes:
\begin{equation}
P(\text{password})=P(S \rightarrow \text{SP}_k) \prod\nolimits_\text{i, j} P(\text{SFT}_i \rightarrow \text{SF}_j).
\label{eq:P_PFSP}
\end{equation}
Researchers have proposed different ways to assign probabilities to \emph{T}. In~\cite{Narayanan-CCS2005}, probabilities of L-segments are calculated by another Markov model over a natural language, while D- and S-segments share the same probability. In~\cite{Weir-SP2009-PCFG}, probabilities of D- and S-segments are calculated based on the training set, while L-segments in a given dictionary are assigned the same probability. Different from the above approaches, we design a more general way to deal with SFs not present in the training set. First, we further split SFs into two sub-sets, observed SFs (marked as OSFs, $T_\text{ob}$) and unobserved SFs (marked as USFs, $T_\text{uob}$). Under these definitions, the probabilities of these two sets are marked as $P_\text{OSFs}$ and $P_\text{USFs}$, respectively. Then we have the following equation: 
\begin{equation}
\forall i, P_\text{OSFs} + P_\text{USFs} = 1, \text{OSFs}, \text{USFs} \in \text{SFT}_i.
\end{equation}
Our smoothing method tries to assign more meaningful probabilities to USFs. In our experiments, we split the training set into two parts according to the size ratio of the training and target databases, then calculate $w_{A,B,i}$ for every $\text{SFT}_i$ following Eq.~\eqref{eq:similarity_SF-SFTi}, and set the estimated probabilities of all OSFs and all USFs under $\text{SFT}_i$ as $P_\text{OSFs} = w_{A,B,i}$ and $P_\text{USFs} = 1 - w_{A,B,i}$. Finally, for each individual SF, we do the following:
\begin{itemize}
\item For an individual $\text{OSF} \in \text{SFT}_i$, its probability is calculated based on its original probability in the training set weighted by $w_{A,B,i}$, i.e., $P(\text{OSF}) = w_{A,B,i} \times P(\text{OSF}|\text{SFT}_i)$.

\item For an individual $\text{USF} \in \text{SFT}_i$, we assume that each USF appears equally, so $P(\text{USF}) = \frac{1 - w_{A,B,i}}{\#(\text{USFs})}$, where $\#(\text{USFs})$ is the number of all USFs in $\text{SFT}_i$.
\end{itemize}
The smoothing method can be easily generalized to handle more complicated cases, e.g., USFs of a specific SFT and different USFs of the same SFT are handled differently from others. The smoothing method on USFs can in principle be generalized to unobserved SPs, too. These will be left as our future work.

\subsection{Experiment Setups}
\label{subsec:experiment_setups}

\textbf{Performance metrics}: To compare the performance of password cracking methods, we need some quantitative metrics. One effective metric widely used in the literature is the ``coverage rate'' $R(\text{D},n)=N_c(\text{D},n)/N(\text{D})\in[0,1]$, where $N(D)$ is the total number of passwords in the target (test) database D and $N_c(D,n)$ is the number of successfully cracked passwords in D with $n$ guesses. In fact, this metric can also be split into two different types:
\begin{itemize}
\item[a)] $R_{\text{po}}(\text{D},n)$. If D has duplicate passwords or password frequencies, this metric can be seen as working at the user-level. The higher $R_{\text{po}}(\text{D},n)$ is, the more users' passwords are cracked.

\item[b)] $R_{\text{pa}}(\text{D},n)$. If D has neither duplicate passwords nor password frequencies, this metric works at the password-level. As reported in~\cite{Chunk-2021-CCS, Pasquini-S&P-2021}, this metric is a good indicator to demonstrate a password cracking method's ability to generating new (or unseen) passwords.
\end{itemize}

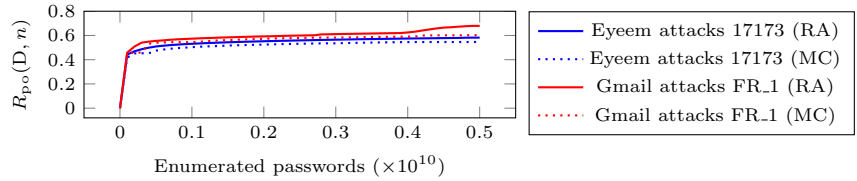
\begin{figure}[!tb]
\centering
\begin{tikzpicture}
\begin{axis}[
width=0.6\linewidth,
height=3cm,
xlabel={Enumerated passwords ($\times 10^{10}$)},
ylabel={$R_{\text{po}}(\text{D},n)$},
x label style={rotate=0,font=\small},
y label style={rotate=0,font=\small},
ymax=0.8,
xtick={0.0, 0.10, 0.20, 0.30, 0.40, 0.50},
x tick label style={rotate=0,font=\small},
y tick label style={rotate=0,font=\small},
legend style={font=\tiny},
legend pos=outer north east
]
\addplot [blue, thick] table {./figures/cracking/data/comparison-mc-ra/data-GMAIL-100W-FR1-NOTRIM-SEMANTIC-RA.txt};
\addplot [blue, dotted, thick] table {./figures/cracking/data/comparison-mc-ra/data-GMAIL-100W-FR1-NOTRIM-SEMANTIC-MC.txt};
\addplot [red, thick] table {./figures/cracking/data/comparison-mc-ra/data-EYEEM-100W-17173-NOTRIM-SEMANTIC-RA.txt};
\addplot [red, dotted, thick] table {./figures/cracking/data/comparison-mc-ra/data-EYEEM-100W-17173-NOTRIM-SEMANTIC-MC.txt};

\legend{Eyeem $\rightarrow$ 17173 (RA), Eyeem $\rightarrow$ 17173 (MC), Gmail $\rightarrow$ FR\_1 (RA), Gmail $\rightarrow$ FR\_1 (MC)}
\end{axis}
\end{tikzpicture}
\caption{Performance using Monte-Carlo (MC) estimation and real-attacks (RA).}
\label{fig:comparison_mc_ra}
\end{figure}

There are two main methods for calculating coverage rates: 1) running a real password cracking process to enumerate passwords and calculate the actual coverage rate, i.e., via a simulated ``real-attacking'', and 2) using a stochastic process like the Monte-Carlo algorithm proposed in~\cite{2015Monte} to approximately estimate the coverage rate. The ``real-attacking'' method can give more accurate results, but can be computationally prohibitive if the number of guessed passwords $n$ becomes too large (e.g., above $10^{12}$). Therefore, when this method is used, it is common to use a practically large but computationally achievable value of $n$, e.g., $n=10^7$~\cite{Wang-USENIX-Security2019, Veras-NDSS2014} and $n=10^{10}$~\cite{Pasquini-S&P-2021}. The Monte-Carlo method can work only with password cracking algorithms based on a clearly defined probability model, but can be used to estimate the coverage rate of a very large $n$ with a much smaller number of randomly sampled passwords (e.g., $10^6$ random passwords to estimate the coverage rate with $n$ as large as $10^{16}$)~\cite{Reasoning-Liu-SP2019, Melicher-USENIX-Security2016, Chunk-2021-CCS}. We chose to use ``real-attacking'' for all our experiments for the following reasons: 1) We hoped to compare our work with as many different models as possible, but \cite{Pasquini-S&P-2021} was clearly claimed that it was not suitable for Monte-Carlo estimation. 2) In~\cite{2015Monte, Melicher-USENIX-Security2016}, it was mentioned that the exact error rates of the Monte-Carlo estimation method depend heavily on the attack methods, so we conducted a small experiment to see whether we could use Monte-Carlo estimation for \CrackerAcronym. Our results in Figure~\ref{fig:comparison_mc_ra} showed that the coverage rates calculated from real-attacking and Monte-Carlo experiments can have a gap as high as 17.79\% for \CrackerAcronym, which we considered too high for a fair and reliable comparison with other SOTA methods. Therefore, to better understand how \CrackerAcronym\ performs, we chose to use ``real-attacking'' metrics for all our experiments. 

\textbf{The SOTA benchmarks}: To investigate how \CrackerAcronym's performance compares against other mainstream SOTA password cracking methods, we used the latest implementation of~\cite{Weir-SP2009-PCFG}, i.e., PCFG ver.\ 4.3 \cite{pcfgcracker} (denoted by ``$\text{PCFG}_w$''), Veras et al.'s Semantic PCFG~\cite{Veras-NDSS2014, semanticcracker} (denoted by ``$\text{PCFG}_{\text{Se}}$''), and Melicher et al.'s neural network~\cite{Melicher-USENIX-Security2016, flacracker} (denoted by ``FLA''), as the benchmarks. We noticed that there are also some other password modeling methods, such as CPG and DPG~\cite{Pasquini-S&P-2021}, PassGAN~\cite{Hitaj-ACNS2019}, RFGuess~\cite{Random-Forest-Wang-US2023}, PassBERT~\cite{Bi-directional-Xu-US2023}, PassTSL~\cite{passtsl-acisp2024}, OMEN~\cite{Durmuth-ESSoS2015}, and one based on an $n$-gram Markov model~\cite{Ma-SP2014}. However, based on the results of~\cite{Pasquini-S&P-2021, Melicher-USENIX-Security2016}, we observed that FLA can always outperform OMEN and 6-gram Markov models. As to RFGuess and PassTSL, we noticed that they evaluated their performance by Monte Carlo estimation. We excluded PassBERT and CPG model as they both require a bunch of preset templates as additional input. The DPG model can be considered an enhanced version of PassGAN, so we conducted some experiments to see how DPG performs without the feedback of target sets using the open-sourced code. We trained DPG using the training sets described in Section~\ref{subsec:experiment_setups} and generated $\times 10^{9}$ passwords in about 3 days. Our results over all testing sets showed that, without the feedback of the testing set, e.g., $\alpha=0$, DPG can outperform PassGAN, but has a much lower performance compared with \CrackerAcronym\ (see Figure~\ref{fig:comparison_dpg}). Considering the above experimental results, as well as the fact that DPG can only obtain character-level semantic information in reality, we finally decided to not consider CPG, DPG, PassGAN, RFGuess, PassBERT, PassTSL or OMEN as part of our benchmarks.

\newcommand\drawlegend[1]{\protect\tikz\protect\draw [thick,#1] plot[] coordinates {(0,0) (0.3,0.1)(0.6,0)};}
\newcommand\drawdottedlegend[1]{\protect\tikz\protect\draw [thick, dotted, #1] plot[] coordinates {(0,0) (0.3,0.1)(0.6,0)};}

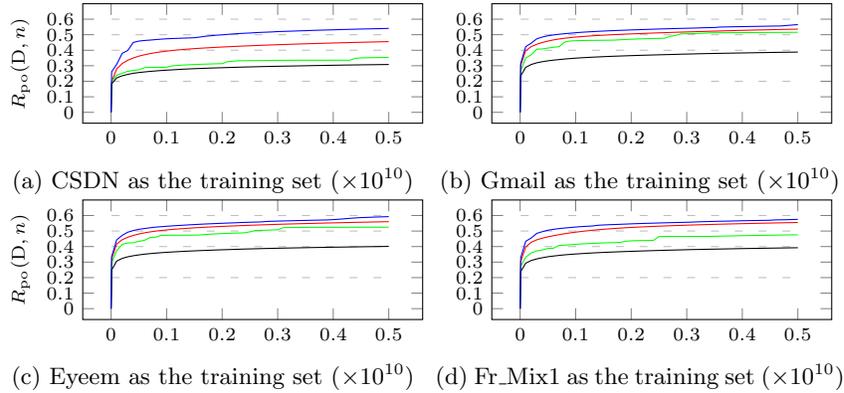
\begin{figure}[!tb]
\centering
\begin{tikzpicture}
\begin{axis}[
width=0.9\linewidth,
height=0.4\linewidth,
xlabel={Enumerated passwords ($\times 10^{9}$) trained with CSDN},
ylabel={$R_{\text{po}}(\text{All},n)$},
x label style={rotate=0,font=\small},
y label style={rotate=0,font=\small},
ymax=0.6,
xtick={0, 0.1, 0.2, 0.3, 0.4, 0.5, 0.6, 0.7, 0.8, 0.9, 1.0},
ytick={0, 0.1, 0.2, 0.3, 0.4, 0.5},
x tick label style={rotate=0,font=\small},
y tick label style={rotate=0,font=\small},
]

\addplot [red] table {./figures/cracking/data/comparison-dpg/appendix-data-CSDN-100W-all-NDPG.txt};
\addplot [green] table {./figures/cracking/data/comparison-dpg/appendix-data-CSDN-100W-all-SEMANTIC.txt};

\end{axis}
\end{tikzpicture}
\caption{Performance comparison between \CrackerAcronym\ and DPG over all testing sets. \drawlegend{green}~\CrackerAcronym, \drawlegend{red}~DPG~\cite{Pasquini-S&P-2021}.}
\label{fig:comparison_dpg}
\end{figure}

\textbf{Training and test sets}: For our experiments, we used CSDN, Gmail, Eyeem, Fr\_Mix1 as training sets, for they have similar sizes and one for each of the four languages studied. The other 13 databases are treated as testing sets, and it ends up to $4\times 13=52$ test cases. More precisely, we used the output of \ModelAcronym\ dealing with each of the four databases as \CrackerAcronym's input, then enumerated a set of passwords to attack each of the other 13 databases under the ``real-attacking'' scene. All these 17 databases have duplicate passwords shared by different users, so that we can investigate the attack performance at two levels: user-level (having duplicated passwords)and password-level (having unique passwords). All the three benchmarking methods and \CrackerAcronym\ are training-based, and we used exactly the same training set to ensure the comparison is fair.

\textbf{Parameter selections}: For all the three benchmarks, we used their default configurations recommended by their authors/developers to generate passwords and calculate guess numbers. Note that the FLA implementation~\cite{flacracker} does not provide a direct interface to generate a specified number of passwords, but can output passwords with their probabilities higher than a given threshold. Therefore, to align with the scale of guessed passwords that previous work used~\cite{Wang-USENIX-Security2019, Veras-NDSS2014, Pasquini-S&P-2021}, we set a threshold of $10^{-12}$ for FLA, which led to maximum $5\times 10^9$ guessed passwords for each training set. This number of guessed passwords is large enough to compare password cracking performance, and to make the computational costs of the experiments manageable in a few weeks\footnote{According to the run-time performance results reported in Section~\ref{subsec:CrackingExperimentalResult}, FLA is the least efficient password generating method. As reported in~\cite{Pasquini-S&P-2021}, FLA would need more than two weeks to generate $10^{10}$ passwords.}.

\subsection{Experimental Results}
\label{subsec:CrackingExperimentalResult}

In this section, we report results of a series of experiments we conducted to show how much our \CrackerAcronym\ benefits from the richer semantic information enabled by \ModelAcronym. We ran all experiments on a machine with an Intel Xeon E5-2640 CPU and two Nvidia Tesla M40 GPUs.

\newlength\sfigwidth
\setlength\sfigwidth{0.5\linewidth}
\newlength\sfigheight
\setlength\sfigheight{0.26\linewidth}
\begin{figure*}[!tb]
\centering
\subfloat[Trained with CSDN ($\times 10^{10}$)]{
	\begin{tikzpicture}
\begin{axis}[
width=0.28\linewidth,
height=0.2\linewidth,
ylabel={$R_{\text{po}}(\text{All},n)$},
y label style={rotate=0,font=\scriptsize},
ymax=0.7,
xtick={0, 0.1, 0.2, 0.3, 0.4, 0.5},
ytick={0, 0.1, 0.2, 0.3, 0.4, 0.5, 0.6},
x tick label style={rotate=0,font=\scriptsize},
y tick label style={rotate=0,font=\scriptsize},
]

\addplot [lightgray, loosely dashed, thin] table {./figures/cracking/data/comparison-population-level/x-axis-1.txt};
\addplot [lightgray, loosely dashed, thin] table {./figures/cracking/data/comparison-population-level/x-axis-2.txt};
\addplot [lightgray, loosely dashed, thin] table {./figures/cracking/data/comparison-population-level/x-axis-3.txt};
\addplot [lightgray, loosely dashed, thin] table {./figures/cracking/data/comparison-population-level/x-axis-4.txt};

\addplot [red] table {./figures/cracking/data/comparison-population-level/data-CSDN-100W-all-FLA.txt};
\addplot [green] table {./figures/cracking/data/comparison-population-level/data-CSDN-100W-all-PCFG.txt};
\addplot [black] table {./figures/cracking/data/comparison-population-level/data-CSDN-100W-all-SEPCFG.txt};
\addplot [blue] table {./figures/cracking/data/comparison-population-level/data-CSDN-100W-all-SEMANTIC.txt};
\end{axis}
\end{tikzpicture}
}
\subfloat[Trained with Gmail ($\times 10^{10}$)]{
	\begin{tikzpicture}
\begin{axis}[
width=0.28\linewidth,
height=0.2\linewidth,
ymax=0.7,
xtick={0, 0.1, 0.2, 0.3, 0.4, 0.5},
ytick={0, 0.1, 0.2, 0.3, 0.4, 0.5, 0.6},
x tick label style={rotate=0,font=\scriptsize},
y tick label style={rotate=0,font=\scriptsize},
]

\addplot [lightgray, loosely dashed, thin] table {./figures/cracking/data/comparison-population-level/x-axis-1.txt};
\addplot [lightgray, loosely dashed, thin] table {./figures/cracking/data/comparison-population-level/x-axis-2.txt};
\addplot [lightgray, loosely dashed, thin] table {./figures/cracking/data/comparison-population-level/x-axis-3.txt};
\addplot [lightgray, loosely dashed, thin] table {./figures/cracking/data/comparison-population-level/x-axis-4.txt};

\addplot [red] table {./figures/cracking/data/comparison-population-level/data-GMAIL-100W-all-FLA.txt};
\addplot [green] table {./figures/cracking/data/comparison-population-level/data-GMAIL-100W-all-PCFG.txt};
\addplot [black] table {./figures/cracking/data/comparison-population-level/data-GMAIL-100W-all-SEPCFG.txt};
\addplot [blue] table {./figures/cracking/data/comparison-population-level/data-GMAIL-100W-all-SEMANTIC.txt};
\end{axis}
\end{tikzpicture}
}
\subfloat[Trained with Eyeem ($\times 10^{10}$)]{
	\begin{tikzpicture}
\begin{axis}[
width=0.28\linewidth,
height=0.2\linewidth,
ymax=0.7,
xtick={0, 0.1, 0.2, 0.3, 0.4, 0.5},
ytick={0, 0.1, 0.2, 0.3, 0.4, 0.5, 0.6},
x tick label style={rotate=0,font=\scriptsize},
y tick label style={rotate=0,font=\scriptsize},
]

\addplot [lightgray, loosely dashed, thin] table {./figures/cracking/data/comparison-population-level/x-axis-1.txt};
\addplot [lightgray, loosely dashed, thin] table {./figures/cracking/data/comparison-population-level/x-axis-2.txt};
\addplot [lightgray, loosely dashed, thin] table {./figures/cracking/data/comparison-population-level/x-axis-3.txt};
\addplot [lightgray, loosely dashed, thin] table {./figures/cracking/data/comparison-population-level/x-axis-4.txt};

\addplot [red] table {./figures/cracking/data/comparison-population-level/data-EYEEM-100W-all-FLA.txt};
\addplot [green] table {./figures/cracking/data/comparison-population-level/data-EYEEM-100W-all-PCFG.txt};
\addplot [black] table {./figures/cracking/data/comparison-population-level/data-EYEEM-100W-all-SEPCFG.txt};
\addplot [blue] table {./figures/cracking/data/comparison-population-level/data-EYEEM-100W-all-SEMANTIC.txt};
\end{axis}
\end{tikzpicture}
}
\subfloat[Trained with Fr\_Mix1 ($\times 10^{10}$)]{
	\begin{tikzpicture}
\begin{axis}[
width=0.28\linewidth,
height=0.2\linewidth,
ymax=0.7,
xtick={0, 0.1, 0.2, 0.3, 0.4, 0.5},
ytick={0, 0.1, 0.2, 0.3, 0.4, 0.5, 0.6},
x tick label style={rotate=0,font=\scriptsize},
y tick label style={rotate=0,font=\scriptsize},
]

\addplot [lightgray, loosely dashed, thin] table {./figures/cracking/data/comparison-population-level/x-axis-1.txt};
\addplot [lightgray, loosely dashed, thin] table {./figures/cracking/data/comparison-population-level/x-axis-2.txt};
\addplot [lightgray, loosely dashed, thin] table {./figures/cracking/data/comparison-population-level/x-axis-3.txt};
\addplot [lightgray, loosely dashed, thin] table {./figures/cracking/data/comparison-population-level/x-axis-4.txt};

\addplot [red] table {./figures/cracking/data/comparison-population-level/data-FR0-100W-all-FLA.txt};
\addplot [green] table {./figures/cracking/data/comparison-population-level/data-FR0-100W-all-PCFG.txt};
\addplot [black] table {./figures/cracking/data/comparison-population-level/data-FR0-100W-all-SEPCFG.txt};
\addplot [blue] table {./figures/cracking/data/comparison-population-level/data-FR0-100W-all-SEMANTIC.txt};
\end{axis}
\end{tikzpicture}
}
\caption{Performance comparison at the user-level between \CrackerAcronym\ and three SOTA password cracking methods over all testing sets on average using real-attacking. \drawlegend{blue}~\CrackerAcronym, \drawlegend{green}~$\text{PCFG}_w$~\cite{pcfgcracker}, \drawlegend{black}~ $\text{PCFG}_{\text{Se}}$~\cite{semanticcracker}, \drawlegend{red}~FLA~\cite{flacracker}.}
\label{fig:performance_on_population_level}
\end{figure*}
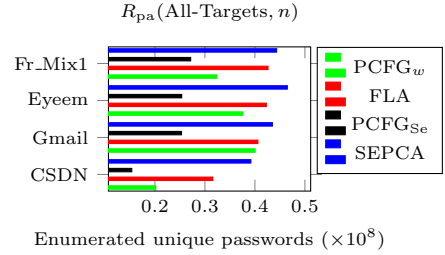

\textbf{Performance Comparison at the User-Level}: Figure~\ref{fig:performance_on_population_level} shows average results of all testing sets at the user-level. There are several clear observations as follows.

In terms of the average performance across 52 test cases, \CrackerAcronym\ performed significantly better than all the benchmarks: it outperformed $\text{PCFG}_w$ by 21.53\%, $\text{PCFG}_{\text{Se}}$ by 52.55\% and FLA by 7.86\%. If we look at all the 52 test cases individually, the results are also overwhelmingly positive: \CrackerAcronym\ outperformed $\text{PCFG}_w$ and $\text{PCFG}_{\text{Se}}$ for all 52 cases, and FLA for all but one case (for the only one the performance drop is negligible at $-0.3\%$). The only slight performance drop when compared with FLA happened when attacking MyHeritage. This exceptional case is not surprising: as mentioned in Section~\ref{subsubsec:CrossDatabaseCorrelations}, users of MyHeritage tended to choose very unique SFTs and SFs, therefore the alignment between the training set and MyHeritage will be poorer. Detailed information are displayed in Table~\ref{tab:detail_population_coverge_rate}. These results indicate that \CrackerAcronym\ can be seen as the most practical and effective method for attacking a given database, as long as the number of guessed password is not prohibitively large (up to the level of $10^{10}$).

\begin{table*}[!htb]
\renewcommand{\arraystretch}{1.2} 
\centering
\scriptsize
\caption{Performance comparison between \CrackerAcronym\ and three state-of-the-art password cracking methods at user-level.}
\label{tab:detail_population_coverge_rate}
\begin{threeparttable}
\begin{tabularx}{\linewidth}{cc*{14}{Y}}
\toprule
& Metrics\tnote{b} & 2 & 3 & 4 & 5 & 6 & 7 & 8 & 9 & 10 & 12 & 14 & 16 & 17 & Average\\
\midrule
\multirow{7}{*}{1} 
& CR($S_1$\tnote{a}) & 74.19\% & 74.06\% & 76.36\% & 74.06\% & 54.33\% & 35.37\% & 17.32\% & 40.27\% & 22.74\% & 31.32\% & 29.46\% & 30.06\% & 32.67\% & 45.56\% \\
& CR($S_2$) & 57.10\% & 54.21\% & 55.42\% & 58.22\% & 39.73\% & 27.68\% & 16.58\% & 32.81\% & 18.33\% & 28.71\% & 23.56\% & 22.32\% & 25.74\% & 35.42\% \\
& CR($S_3$) & 42.59\% & 40.88\% & 43.10\% & 47.49\% & 31.67\% & 23.79\% & 17.00\% & 34.11\% & 18.70\% & 30.92\% & 23.84\% & 22.52\% & 24.02\% & 30.82\% \\
& CR($S_4$) & \textbf{74.62\%}\tnote{c} & \textbf{74.80\%} & \textbf{76.78\%} & \textbf{76.59\%} & \textbf{55.84\%} & \textbf{48.90\%} & \textbf{23.43\%} & \textbf{60.00\%} & \textbf{33.08\%} & \textbf{43.69\%} & \textbf{42.53\%} & \textbf{45.14\%} & \textbf{47.35\%} & \textbf{54.06\%} \\
\cline{2-16}
& RIR($S_1$) & 0.58\% & 1.00\% & 0.54\% & 3.42\% & 2.79\% & 38.25\% & 35.28\% & 48.99\% & 45.48\% & 39.48\% & 44.35\% & 50.16\% & 44.92\% & 27.33\% \\
& RIR($S_2$) & 30.67\% & 37.97\% & 38.53\% & 31.56\% & 40.56\% & 76.68\% & 41.36\% & 82.87\% & 80.50\% & 52.15\% & 80.50\% & 102.2\% & 83.94\% & 59.96\% \\
& RIR($S_3$) & 75.18\% & 82.99\% & 78.15\% & 61.29\% & 76.31\% & 105.5\% & 37.89\% & 75.90\% & 76.97\% & 41.30\% & 78.42\% & 100.4\% & 97.14\% & 75.96\% \\
\hline
\multirow{7}{*}{11} 
& CR($S_1$) & 65.99\% & 62.86\% & 58.53\% & 58.98\% & 32.67\% & 56.72\% & 31.34\% & 70.18\% & \textbf{41.36\%} & 54.79\% & 50.51\% & 55.55\% & 57.41\% & 53.61\% \\
& CR($S_2$) & 63.58\% & 59.20\% & 55.31\% & 56.63\% & 29.99\% & 55.70\% & 30.74\% & 69.23\% & 40.93\% & 52.52\% & 47.93\% & 52.56\% & 55.94\% & 51.56\% \\
& CR($S_3$) & 27.76\% & 25.74\% & 27.53\% & 32.88\% & 17.83\% & 45.91\% & 28.98\% & 63.04\% & 35.67\% & 58.70\% & 45.69\% & 46.95\% & 47.88\% & 38.81\% \\
& CR($S_4$) & \textbf{67.18\%} & \textbf{64.96\%} & \textbf{61.69\%} & \textbf{61.96\%} & \textbf{35.77\%} & \textbf{59.91\%} & \textbf{33.72\%} & \textbf{73.52\%} & 41.23\% & \textbf{61.99\%} & \textbf{54.40\%} & \textbf{58.22\%} & \textbf{60.41\%} & \textbf{56.54\%} \\
\cline{2-16}
& RIR($S_1$) & 1.82\% & 3.34\% & 5.39\% &  5.04\% &  9.47\% &  5.63\% &  7.60\% &  4.76\% &  -0.3\% &  13.14\% &  7.70\% &  4.81\% &  5.22\% &  5.66\% \\
& RIR($S_2$) & 5.67\% & 9.72\% & 11.52\% & 9.42\% & 19.26\% & 7.55\% & 9.68\% & 6.20\% & 0.75\% & 18.03\% & 13.49\% & 10.77\% & 7.98\% & 10.00\% \\
& RIR($S_3$) & 141.9\% & 152.3\% & 124.1\% & 88.46\% & 100.5\% & 30.48\% & 16.35\% & 16.62\% & 15.59\% & 5.60\% & 19.07\% & 24.01\% & 26.16\% & 58.56\% \\
\hline
\multirow{7}{*}{13} 
& CR($S_1$) & 67.30\% & 65.08\% & 62.78\% & 63.02\% & 37.99\% & 57.37\% & 33.13\% & 70.02\% & 40.13\% & 64.59\% & 53.04\% & 56.73\% & 57.28\% & 56.04\% \\
& CR($S_2$) & 60.66\% & 55.20\% & 52.73\% & 54.39\% & 30.93\% & 56.24\% & 33.09\% & 69.51\% & 39.81\% & 65.18\% & 52.06\% & 55.33\% & 56.86\% & 52.46\% \\
& CR($S_3$) & 30.29\% & 28.52\% & 31.18\% & 35.31\% & 20.70\% & 45.31\% & 29.81\% & 62.89\% & 34.52\% & 60.34\% & 46.41\% & 48.13\% & 47.96\% & 40.10\% \\
& CR($S_4$) & \textbf{69.95\%} & \textbf{68.65\%} & \textbf{67.96\%} & \textbf{67.85\%} & \textbf{42.48\%} & \textbf{60.92\%} & \textbf{34.65\%} & \textbf{74.47\%} & \textbf{41.00\%} & \textbf{66.97\%} & \textbf{55.82\%} & \textbf{58.84\%} & \textbf{60.67\%} & \textbf{59.25\%} \\
\cline{2-16}
& RIR($S_1$) & 3.94\% &  5.48\% &  8.27\% &  7.66\% &  11.80\% &  6.18\% &  4.58\% &  6.35\% &  2.17\% &  3.69\% &  5.25\% &  3.73\% &  5.92\% &  5.77\% \\
& RIR($S_2$) & 15.32\% & 24.36\% & 28.88\% & 24.74\% & 37.32\% & 8.32\% & 4.73\% & 7.13\% & 3.00\% & 2.73\% & 7.23\% & 6.35\% & 6.69\% & 13.60\% \\
& RIR($S_3$) & 130.9\% & 140.7\% & 117.9\% & 92.17\% & 105.2\% & 34.45\% & 16.23\% & 18.40\% & 18.77\% & 10.98\% & 20.30\% & 22.26\% & 26.51\% & 58.07\% \\
\hline
\multirow{7}{*}{15} 
& CR($S_1$) & 65.34\% & 62.65\% & 59.46\% & 60.03\% & 36.01\% & 59.79\% & 33.61\% & 68.99\% & 40.86\% & 63.77\% & 54.03\% & 58.55\% & 58.48\% & 55.51\% \\
& CR($S_2$) & 49.50\% & 43.58\% & 41.95\% & 45.51\% & 25.74\% & 54.58\% & 32.49\% & 64.38\% & 38.61\% & 61.29\% & 50.67\% & 55.00\% & 55.14\% & 47.57\% \\
& CR($S_3$) & 27.41\% & 25.45\% & 27.57\% & 33.09\% & 18.50\% & 46.86\% & 29.67\% & 60.74\% & 36.76\% & 58.86\% & 46.84\% & 49.50\% & 48.97\% & 39.25\% \\
& CR($S_4$) & \textbf{66.33\%} & \textbf{63.56\%} & \textbf{60.76\%} & \textbf{61.69\%} & \textbf{36.92\%} & \textbf{62.32\%} & \textbf{35.20\%} & \textbf{73.26\%} & \textbf{41.65\%} & \textbf{65.95\%} & \textbf{56.65\%} & \textbf{60.52\%} & \textbf{61.81\%} & \textbf{57.43\%} \\
\cline{2-16}
& RIR($S_1$) & 1.51\% &  1.45\% &  2.18\% &  2.77\% &  2.53\% &  4.24\% &  4.76\% &  6.18\% &  1.94\% &  3.42\% &  4.85\% &  3.37\% &  5.70\% &  3.45\% \\
& RIR($S_2$) & 33.98\% & 45.85\% & 44.84\% & 35.55\% & 43.47\% & 14.18\% & 8.36\% & 13.78\% & 7.89\% & 7.59\% & 11.81\% & 10.04\% & 12.10\% & 22.26\% \\
& RIR($S_3$) & 141.9\% & 149.7\% & 120.4\% & 86.46\% & 99.58\% & 33.00\% & 18.65\% & 20.61\% & 13.29\% & 12.04\% & 20.94\% & 22.28\% & 26.22\% & 58.86\% \\
\bottomrule
\end{tabularx}
\begin{tablenotes}
\item[a] Denotations of cracking methods: $S_1$ -- FLA~\cite{flacracker}, $S_2$ -- $\text{PCFG}_w$~\cite{pcfgcracker}, $S_3$ -- $\text{PCFG}_{\text{Se}}$~\cite{semanticcracker}, $S_4$ -- \CrackerAcronym. All experiments are conducted across $4\times 13=52$ different test cases (4 training databases in the first column and 13 target databases in Columns 3 to 15) at user-level. 

\item[b] Performance metrics in Column 2: CR = Coverage Rate, RIR = Relative Improvement Rate defined as $\text{RIR}(x) =(S_4-x)/x$. 

\item[c] The items in bold indicate that the method has the best cracking performance on the corresponding target database.

\end{tablenotes}
\end{threeparttable}
\end{table*}

\begin{table}[!htb]
\renewcommand{\arraystretch}{1.4}
\centering
\scriptsize
\caption{Comparison between \CrackerAcronym\ and three state-of-the-art methods on password-level over all targets.}
\label{tab:performance_on_password_level}
\begin{threeparttable}
\begin{tabularx}{\linewidth}{lccccc}

\toprule
\multirow{2}{*}{\textbf{Methods}} & \multicolumn{4}{c}{\textbf{Training Sets}} & \multirow{2}{*}{\textbf{AIR (\%)\tnote{a}}} \\
\cmidrule(lr){2-5}
& 1 & 11 & 13 & 15 &  \\
\midrule
FLA~\cite{flacracker} & 20182435 & 26690266 & 27731271 & 28383124 & 11.16\% \\
PCFG$_w$~\cite{pcfgcracker} & 12584740 & 26172639 & 23947390 & 21048687 & 43.83\% \\
PCFG$_{\text{Se}}$~\cite{semanticcracker} & 9430673 & 16872607 & 16713065 & 18518927 & 94.11\% \\
\CrackerAcronym & \textbf{25148844}\tnote{b} & \textbf{28404466} & \textbf{30711293} & \textbf{29198645} & - \\
\bottomrule

\end{tabularx}
\begin{tablenotes}

\item[a] AIR = Average Improvement Rate. Improvement Rate (IR) is defined as IR(x) = (\CrackerAcronym-x)/x.

\item[b] The items in bold indicate that the method has the most unique passwords cracked over all target databases .

\end{tablenotes}
\end{threeparttable}
\end{table}

\textbf{Performance at (Unique) Password-Level}: As shown in Table~\ref{tab:performance_on_password_level}, \CrackerAcronym\ outperformed the benchmarks significantly on all 52 test cases: $\text{PCFG}_w$ by 43.83\%, $\text{PCFG}_{\text{Se}}$ by 94.11\%, and FLA by 11.16\%. In terms of individual test cases, \CrackerAcronym\ performed the best in 50 out of all 52 cases (96.15\%), except for using Gmail, Eyeem and Fr\_Mix1 to attack MyHeritage. Again, as mentioned before, the poorer results on MyHeritage is not surprising given the database-correlation results in Section~\ref{subsubsec:CrossDatabaseCorrelations}.

\begin{table}[!htb]
\centering
\caption{Average coverage rate on user-level aligned to the language.}
\label{tab:performance_in_language_angle}
\begin{tabular}{*{5}{c}}
\toprule
Training Sets & 1 & 11 & 13 & 15\\
\midrule
CN & 71.73\% & 58.31\% & 63.38\% & 57.85\%\\
EN & 41.36\% & 52.10\% & 52.76\% & 53.10\%\\
GE & 43.10\% & 58.19\% & 61.39\% & 61.30\%\\
FR & 46.24\% & 59.32\% & 59.76\% & 61.17\%\\
\bottomrule
\end{tabular}
\end{table}

\textbf{Performance on different language settings}: Table~\ref{tab:performance_in_language_angle} shows a number of interesting observations, which also echo some visual patterns in Figure~\ref{fig:Correlation_of_SFs_and_SFTs} in Section~\ref{subsec:PasswordSemanicsResults}. 1) For Chinese, German and French targets, \CrackerAcronym\ performed better when being trained using language-aligned settings. 2) For English targets, training using an English database does not always produce the best results (52.10\% in Gmail attacking English databases, lower than 53.10\% in Fr\_Mix1), which indicates that English databases likely include users with more diverse backgrounds. Actually this phenomenon is not surprising since Fr\_Mix1 has a higher correlation with English databases than Gmail (Fr\_Mix1, 93.29\% vs.\ Gmail, 91.53\% on SFT-level, while 74.93\% vs.\ 72.18\% on SF-level). 3) The performances of \CrackerAcronym\ are more robust compared to other benchmarks: no matter which database was used for training, \CrackerAcronym\ always have a similar performance to attack all test sets (CSDN: 54.06\%, Gmail: 56.54\%, Eyeem: 59.25\%, Fr\_Mix1: 57.43\%), while $\text{PCFG}_w$ fluctuates in the range from 35.42\% to 52.46\%, FLA moves between 45.56\% and 56.04\%, and $\text{PCFG}_{\text{Se}}$ from 30.81\% to 40.10\%.

\textbf{Run-time performance}:
Table~\ref{tab:average_speed_generating_passwords} shows the run-time performance of  the password generation process of \CrackerAcronym\ and each of the three benchmarks. One can see that \CrackerAcronym\ is much faster (around 5.6 times) on password generation than FLA, although is slower than the other two benchmarks -- around 4.4 times slower than $\text{PCFG}_w$ and around 2.6 times slower than $\text{PCFG}_{\text{Se}}$ (likely due to its utilization of richer semantics). Considering \CrackerAcronym\ outperformed other benchmarks in its password cracking performance, we consider the effectiveness-efficiency balance of \CrackerAcronym\ reasonable.

\begin{table}[!htb]
\centering
\caption{The average speed of generating passwords (p/s = passwords per second).}
\label{tab:average_speed_generating_passwords}
\begin{tabular}{*{4}{c}}
\toprule
SEPCA & $\text{PCFG}_w$~\cite{pcfgcracker} &  $\text{PCFG}_{\text{Se}}$~\cite{semanticcracker} & FLA~\cite{flacracker}\\
\midrule
32,258 p/s & 140,880 p/s & 82,595 p/s & 5,787 p/s\\
\bottomrule
\end{tabular}
\end{table}

\section{Further Discussions}
\label{sec:further_discussions}

The enhanced semantic analysis power of \ModelAcronym\ and the improved password cracking capabilities of \CrackerAcronym\ have many profound implications in real-world applications. In addition to providing researchers with new tools for studying password security and usability, end users of password systems can also benefit from our work, e.g., they have better insights on how to define stronger but still usable passwords, and cyber security professionals have more evidence on how to define password policies to enforce or nudge securer password creation behaviors.

To demonstrate how our experimental results can help inform end users about weak passwords, in Table~\ref{tab:MostCrackedSPs} we list top 10 weakest (i.e., the easiest to crack) password semantic patterns (SPs) for each of the four subsets of password databases grouped by language, leading to in total 20 SPs representing different types of weak passwords. As can be seen from this table, users speaking different languages have different weak password behaviors, but there are also shared patterns such as the use of numbers, dates, names of different types, nouns, and Pinyin for Chinese names. The behavioral patterns have clear psychological reasons since people tend to use things they can remember to define passwords. Such behaviors can be changed by introducing stricter password policies and adopting more intelligent password checkers, and the methods and tools reported in this paper can be used to continuously monitor leaked passwords and to support pentesting exercises simulating password cracking activities of adversarial actors.

\begin{table*}[!tb]
\centering
\caption{Top 10 cracked SPs for each of the four subsets of password databases grouped by language, leading to in total 20 SPs representing different types of weak passwords.}
\label{tab:MostCrackedSPs}
\begin{tabular}{*{5}{c}}
\toprule
SP & CN (\%) & EN (\%) & GE (\%) & FR (\%)\\
\midrule
{[\textsc{NUMBER6}]} & \textbf{12.0}\tnote{a} & \textbf{3.0} & \textbf{2.4} & \textbf{3.2}\\
{[\textsc{NUMBER7}]} & \textbf{10.1} & \textbf{1.4} & 0.9 & \textbf{1.5}\\
{[\textsc{NUMBER8}]} & \textbf{6.3} & \textbf{1.5} & \textbf{1.6} & \textbf{2.0}\\
{[\textsc{NUMBER9}]} & \textbf{3.8} & 0.5 & 0.6 & 0.7\\
{[\textsc{YYMMDD}]} & \textbf{3.4} & 0.4 & 0.2 & 0.3\\
{[\textsc{YYYYMMDD}]} & \textbf{3.0} & $<$ 0.1 & $<$ 0.1 & $<$ 0.1\\
{[\textsc{WKNE}]} & \textbf{1.3} & \textbf{3.8} & \textbf{4.6} & \textbf{5.2}\\
{[\textsc{PY}, \textsc{PY}, \textsc{PY}]} & \textbf{1.1} & 0.2 & 0.2 & $<$ 0.1\\
{[\textsc{MMDDYY}]} & \textbf{1.0} & 0.8 & 0.4 & 1.0\\
{[\textsc{PRE1}, \textsc{NUMBER7}]} & \textbf{0.9} & $<$ 0.1 & $<$ 0.1 & $<$ 0.1\\
{[\textsc{FEMALE\_NAME}]} & $<$ 0.1 & \textbf{3.4} & \textbf{2.8} & \textbf{4.8}\\
{[\textsc{EN\_NOUN}]} & 0.2 & \textbf{2.8} & \textbf{3.0} & \textbf{2.9}\\
{[\textsc{FEMALE\_NAME}, \textsc{NUMBER2}]} & $<$ 0.1 & \textbf{2.1} & \textbf{3.9} & \textbf{2.0}\\
{[\textsc{EN\_NOUN}, \textsc{NUMBER2}]} & $<$ 0.1 & \textbf{1.3} & \textbf{2.0} & 0.9\\
{[\textsc{WKNE}, \textsc{NUMBER2}]} & 0.3 & \textbf{1.3} & \textbf{1.8} & 1.4\\
{[\textsc{FEMALE\_NAME}, \textsc{SUF1}]} & $<$ 0.1 & \textbf{1.3} & $<$ 0.1 & \textbf{1.5}\\
{[\textsc{FEMALE\_NAME}, \textsc{NUMBER3}]} & 0.3 & 0.9 & \textbf{1.5} & 0.7\\
{[\textsc{FEMALE\_NAME}, \textsc{YEAR}]} & $<$ 0.1 & 0.6 & \textbf{1.4} & 0.5\\
{[\textsc{MALE\_NAME}]} & $<$ 0.1 & 1.0 & 1.0 & \textbf{1.6}\\
{[\textsc{DDMMYY}]} & 0.2 & 0.7 & 0.5 & \textbf{1.5}\\
\bottomrule
\end{tabular}
\end{table*}

Based on observed weak passwords, we can derive tips that can help non-expert users to define stronger passwords. For instance, if we use the weak passwords shown in Table~\ref{tab:MostCrackedSPs} as examples, we can provide the following suggestions to end users: 1) avoid using personally identifiable information (e.g., one's own or family members' names and birthdays), commonly used names and other nouns in different languages including Pinyin names in Chinese; 2) rather than using any semantic factors directly, transforming or obfuscating them using different methods to make them harder to guess; 3) constructing passwords using three or more different semantic factors to increase the semantic complexity; and 4) using random passwords with a password manager whenever possible. Note that simply adding prefixes or suffixes before or after a single semantic factor does not effectively increase password security, as two such password patterns are among the weak patterns shown in Table~\ref{tab:MostCrackedSPs}. The user-facing suggestions should not be taken rigidly and statically, since human users' password composition behaviors and password cracking techniques are both constantly evolving.


\section{Ethical Considerations}
\label{section:Ethics}

We considered ethical issues in our research following the common practice followed by other researchers. We used only password databases that were already leaked publicly, many of which are widely used standard databases in password-related research in the literature. We removed all non-password personal information from the databases and kept only passwords themselves for our research. We did not and will not redistribute the password databases we used to avoid potential misuse. Instead, reproducibility is supported by providing sufficient details of the password databases used and how we processed them.

\section{Conclusion}
\label{sec:Conclusion}

This paper presents \ModelAcronym, a new framework and an associated computational process for analyzing password semantics in four different levels. By applying it to 17 leaked databases, we demonstrated how the framework can be used to produce useful new insights about password semantics and the underlying user behaviors. Then, we further proposed \CrackerAcronym, a semantic-aware password cracking architecture equipped by a general smoothing method. Our experiments with the 17 leaked databases showed that \CrackerAcronym\ could outperform other SOTA password cracking methods and it also performed very robustly across 52 test cases with different pairs of training and testing databases.






\bibliographystyle{IEEEtran}
\bibliography{main.bib}

\vskip -2\baselineskip plus -1fil

\begin{IEEEbiography}[{\includegraphics[width=1in,height=1.25in,clip,keepaspectratio]{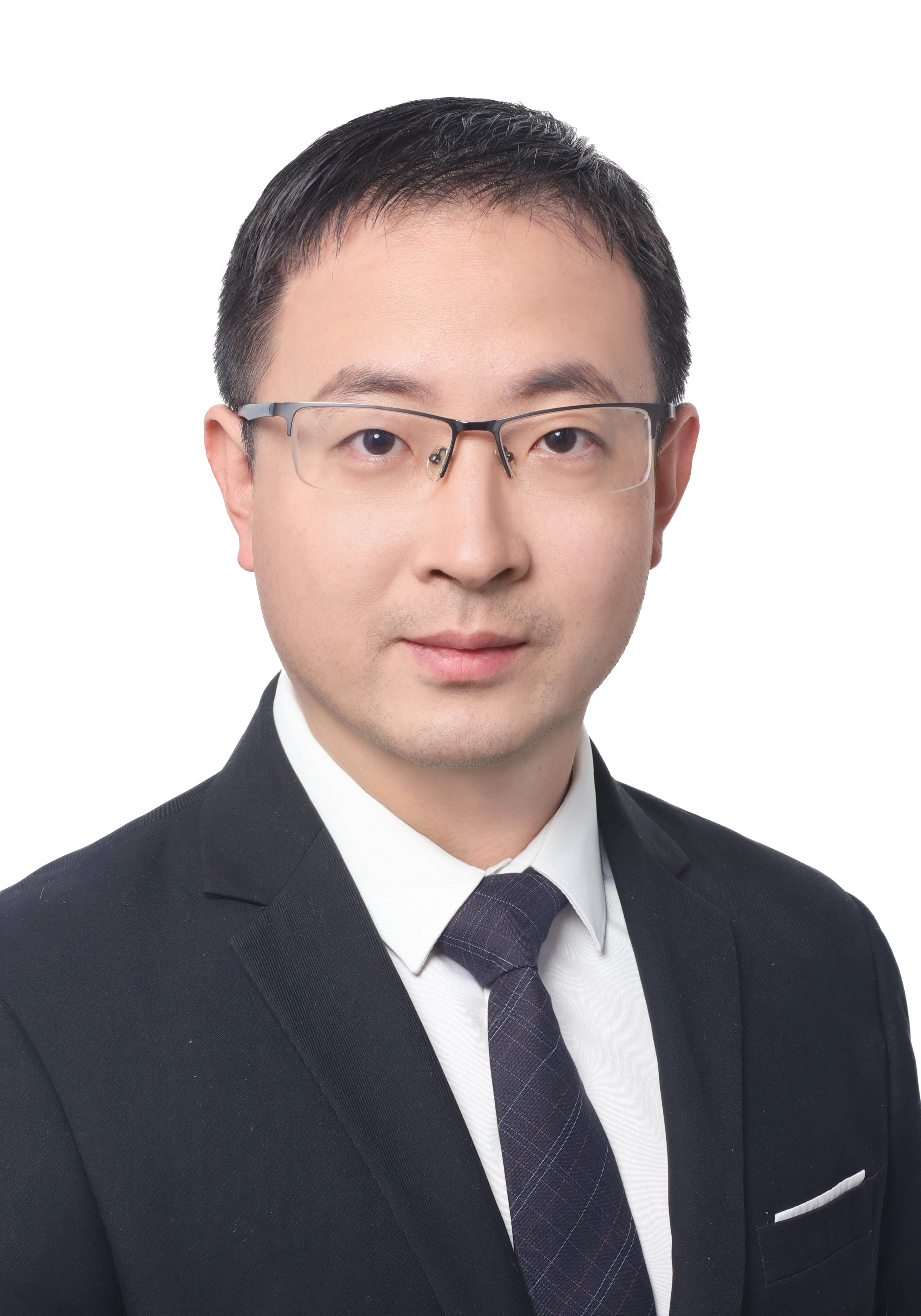}}]{Yangde Wang} received the M.S.\ degree in Information Security from Shanghai Jiao Tong University, Shanghai, China, in 2013. He is currently a Ph.D.\ candidate in the School of Cyber Science and Engineering, Shanghai Jiao Tong University. His main research areas include computer forensics and password security.
\end{IEEEbiography}

\vskip -2\baselineskip plus -1fil

\begin{IEEEbiography}[{\includegraphics[width=1in,height=1.25in,clip,keepaspectratio]{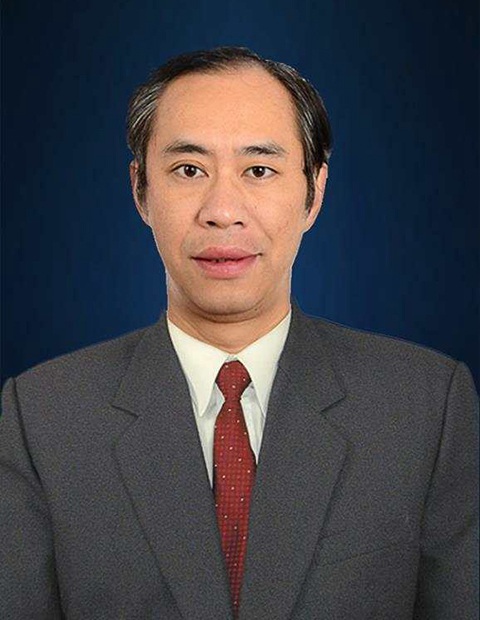}}]{Weidong Qiu} received the Ph.D.\ degree in Computer Software Theory from Shanghai Jiao Tong University, Shanghai, China, in 2001 and received the M.S.\ degree in Cryptography from Xidian University, Xi'an, China, in 1998. He is currently a professor and doctoral supervisor in the School of Cyber Science and Engineering, Shanghai Jiao Tong University. His main research areas include data science, privacy computing, cryptography and computer forensics.
\end{IEEEbiography}

\vskip -2\baselineskip plus -1fil

\begin{IEEEbiography}[{{\includegraphics[width=1in,height=1.25in,clip,keepaspectratio]{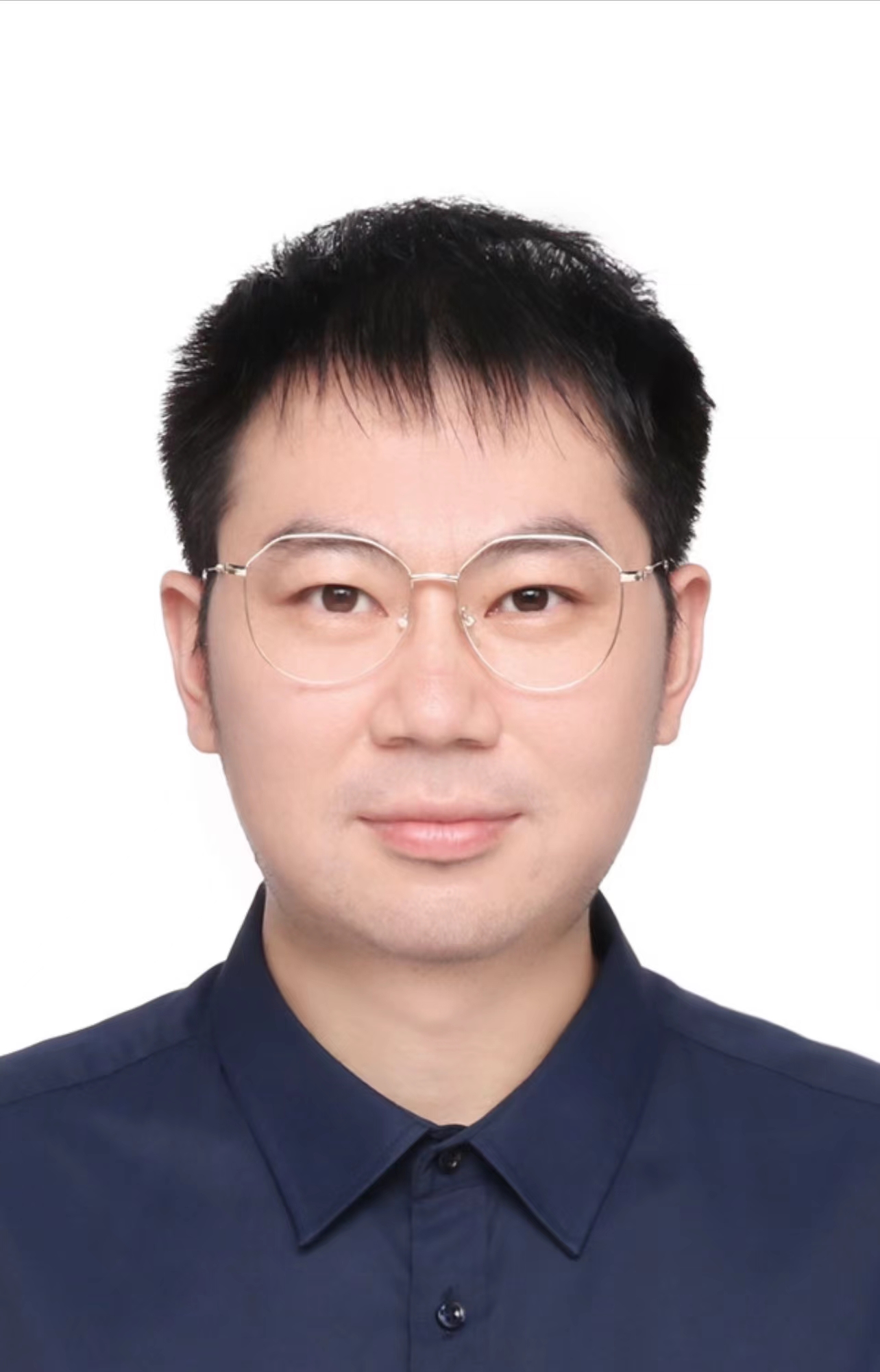}}}]{Peng Tang} is a postdoctoral in the School of Cyber Science and Engineering, at Shanghai Jiao Tong University. He received his M.S.\ degree in Computer Science from Beijing University of Posts and Telecommunications in 2017 and his Ph.D.\ degree in Cyber Security from Shanghai Jiao Tong University in 2022. His research focuses on privacy protection, data science, and AI security.
\end{IEEEbiography}

\vskip -2\baselineskip plus -1fil

\begin{IEEEbiography}[{\includegraphics[width=1in,height=1.25in,clip,keepaspectratio]{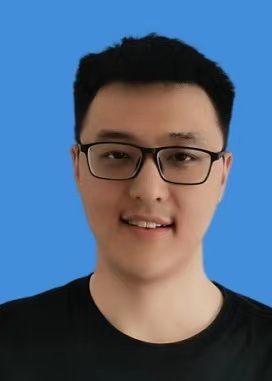}}]{Hao Tian} received the M.S.\ degree in Electronics and Communications Engineering from Shanghai Jiao Tong University in 2021. He is currently working at Haitong Securities, with research interests that include AI security and data circulation security.
\end{IEEEbiography}

\vskip -2\baselineskip plus -1fil

\begin{IEEEbiography}[{\includegraphics[width=1in,height=1.25in,clip,keepaspectratio]{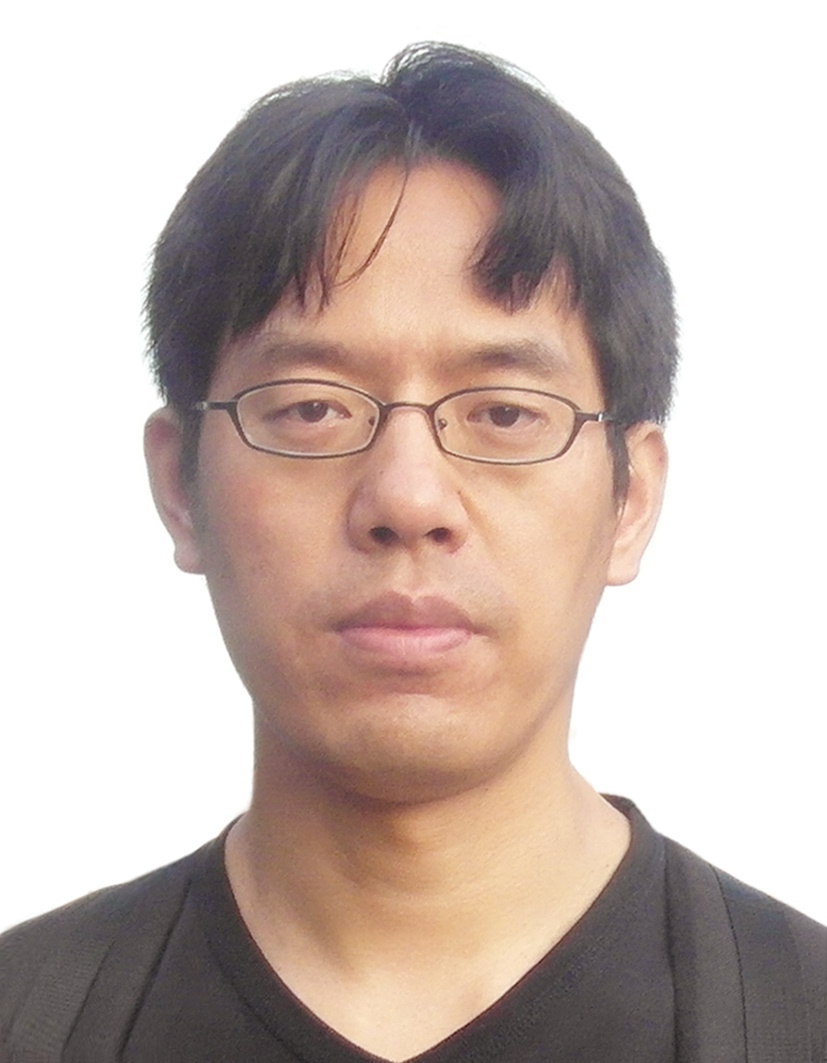}}]{Shujun Li} (M'2008, SM'2012) received his B.E.\ degree in Information Science and Engineering and Ph.D.\ degree in Information and Communication Engineering, both from Xi'an Jiaotong University, China, in 1997 and 2003, respectively. He is Professor of Cyber Security at the School of Computing and Director of the Institute of Cyber Security for Society (iCSS), University of Kent, U.K. His research interests are mostly about inter-disciplinary topics related to cyber security and privacy, human factors, digital forensics and cybercrime, multimedia computing, AI and data science. His work covers multiple application domains, including but not limited to cybercrime, social media analytics, digital health, smart cities, smart homes, and e-tourism. He received multiple awards and honors including the 2022 IEEE Transactions on Circuits and Systems Guillemin-Cauer Best Paper Award.
\end{IEEEbiography}

\vfill
\clearpage

\end{document}